%% file: main.tex
\documentclass{article}

\usepackage[nonatbib,preprint]{neurips_2022}

\usepackage[utf8]{inputenc} 
\usepackage[T1]{fontenc}    
\usepackage{hyperref}       
\usepackage{url}            
\usepackage{booktabs}       
\usepackage{amsfonts}       
\usepackage{nicefrac}       
\usepackage{microtype}      
\usepackage{xcolor}         
\usepackage{tabto}    
\usepackage[superscript,biblabel]{cite}
\usepackage{setspace}  
\usepackage{titlesec}
\usepackage{gensymb}

\usepackage{subfig}
\usepackage{wrapfig}
\usepackage{tikz}
\usepackage{tikz-cd}
\usetikzlibrary{cd}
\usepackage{microtype}
\usepackage{graphicx}
\usepackage{booktabs} 
\usepackage{pifont}       
\usepackage{bbding}       %
\usepackage{fontawesome}  
\usepackage{jabbrv}

\usepackage{amsmath}
\usepackage{amssymb}
\usepackage{mathtools}
\usepackage{amsthm}
\usepackage{arydshln}
\usepackage[capitalize,noabbrev]{cleveref}

\usepackage{mathtools} 
\input{math.tex}

\makeatletter
\renewcommand{\fnum@figure}{\textbf{Figure \thefigure}. }
\renewcommand{\fnum@table}{\textbf{Table \thetable \ |}}

\makeatother
\usepackage[labelsep=space]{caption}
\usepackage[
singlelinecheck=false 
]{caption}

\usepackage{authblk}

\title{
Building-Block Aware Generative Modeling for 3D Crystals of Metal Organic Frameworks 
}

\author[1, 2, $ \dag $, *]{Chenru Duan}
\author[3, $ \dag $, *]{Aditya Nandy}
\author[4, 5, $ \dag $]{Sizhan Liu}
\author[6]{Yuanqi Du}
\author[4, 5]{Liu He}
\author[7]{Yi Qu}
\author[1]{Haojun Jia}
\author[4, 5 *]{Jin-Hu Dou}
\affil[1]{Deep Principle, Inc., Cambridge, MA, 02139}
\affil[2]{Frontiers Science Center for Transformative
Molecules, School of Artificial Intelligence, Shanghai Jiao Tong University, Shanghai 200240, China}
\affil[3]{Department of Chemical and Biomolecular Engineering, University of California, Los Angeles,
Los Angeles, CA 90095}
\affil[4]{National Key Laboratory of Advanced Micro and Nano Manufacture Technology, School of Materials Science and Engineering, Peking University, Beijing
 100871, China. }
\affil[5]{Key Laboratory of Polymer Chemistry and Physics of Ministry of Education, School of Materials Science and Engineering, Peking University,
 Beijing 100871, China.}
\affil[6]{Department of Computer Science, Cornell University, Ithaca, NY, 14850}
\affil[7]{Harvard Law School, Cambridge, MA, USA}
\affil[$ \dag $]{These authors contribute equally}
\affil[*]{Correspondence to: duanchenru@gmail.com, aditya.nandy@ucla.edu, doujinhu@pku.edu.cn}

\begin{document}

\maketitle

\begin{abstract}
Metal–organic frameworks (MOFs) marry inorganic nodes, organic edges, and topological nets into programmable porous crystals, yet their astronomical design space defies brute-force synthesis. 
Generative modeling holds ultimate promise, but existing models either recycle known building blocks or are restricted to small unit cells. 
We introduce Building-Block-Aware MOF Diffusion (BBA MOF Diffusion), an SE(3)-equivariant diffusion model that learns 3D all-atom representations of individual building blocks, encoding crystallographic topological nets explicitly. 
Trained on the CoRE-MOF database, BBA MOF Diffusion readily samples MOFs with unit cells containing 1000 atoms with great geometric validity, novelty, and diversity mirroring experimental databases. 
Its native building-block representation produces unprecedented metal nodes and organic edges, expanding accessible chemical space by orders of magnitude.
One high-scoring $\mathrm{[Zn(1,4-TDC)(EtOH)_{2}]}$ MOF predicted by the model was synthesized, where powder X-ray diffraction, thermogravimetric analysis, and N2 sorption confirm its structural fidelity.
BBA-Diff thus furnishes a practical pathway to synthesizable and high-performing MOFs.
\end{abstract}

\setstretch{1.8}

\input{01_introduction}

\input{02_results}

\input{03_discussion}
\input{04_methods}

\input{05_others}

\bibliographystyle{naturemag_doi}
\bibliography{main.bib}

\clearpage

\setstretch{1}

\input{09_appendix}

\end{document}

%% file: math.tex
\usepackage{amsmath,amsfonts,bm}

\def\1{\bm{1}}

\DeclareMathAlphabet{\mathsfit}{\encodingdefault}{\sfdefault}{m}{sl}
\SetMathAlphabet{\mathsfit}{bold}{\encodingdefault}{\sfdefault}{bx}{n}



%% file: 01_introduction.tex
\section*{1. Introduction}
\qquad Metal-organic frameworks (MOFs) are a type of functional inorganic-organic hybrid material. These  materials connect inorganic metal centers (metal ions or metal-oxyhydroxy clusters) through coordination bonds to form an infinite, periodic extension of the network structure. Due to their unique, modular structure and performance advantages, they have received a great deal of attention in recent years.\cite{stein1993turning,yaghi1998synthetic,yaghi2000design,yaghi2003reticular,greed2025man,dou2021atomically,murray2009hydrogen}
To begin, a highly ordered and programmable aperture network is constructed due to its special coordination mode through metal nodes and organic ligands. Its pore size can be precisely\cite{Forse2020} regulated by changing the ligand length and the size of the metal cluster. 
\cite{deng2010multiple,eddaoudi2002systematic,lu2014tuning}
Secondarily, MOFs exhibit remarkably high specific surface areas, significantly exceeding those of conventional inorganic materials (e.g. zeolites) due to their fully penetrating three-dimensional channels and the efficient use of space by the hybrid inorganic-organic frame.
\cite{furukawa2013chemistry,chae2004route}
More importantly, the dynamic tunability arising from the metal-ligand coordination endows MOFs with exceptional multifunctionality. This characteristic enables nearly unlimited modular combination strategies through rational selection of building blocks.  Specifically, metal nodes serve dual functions as both structural anchors and catalytically active centers, while organic ligands allow programmable modification of pore characteristics through precise functional group engineering.
\cite{baumann2019metal,yuan2018stable}
This "structure-function integration" design strategy allows MOFs to show irreplaceable advantages in the field of intelligent response materials and precise molecular recognition.
Its unique and excellent internal structure determines its position as a core material for a wide range of applications in interdisciplinary fields. MOF materials have a wide range of applications in the fields of energy,\cite{sheberla2017conductive,niu2022conductive,yang2024oxygen} catalysis,\cite{stanley2024analysis,fu2012amine,niu2019promoting,niu2018metal} sensing,\cite{kreno2012metal,allendorf2009luminescent,jin2024metal} gas storage,\cite{li2011carbon,murray2009hydrogen} and so on.

\qquad Over more than two decades of study, experimental structures of MOFs have been collected, sanitized, and curated in the Cambridge Structural Database (CSD).\cite{CSDGroom} 
Only a subset of MOFs are porous; accordingly, different sets of experimentally accessed MOFs have been further sanitized and chosen for usable databases such as CoRE MOF\cite{Chung2019,Zhao2024_CoRE_MOF_DB} and CSD MOF.\cite{CSDMOFFairen2017}
Due to their modular, building-block nature for construction, MOFs can also be assembled from their constituent parts. For this reason, many works over the past decade have constructed hypothetical databases of MOFs with varying degrees of diversity and detail.\cite{Wilmer2012, BoydWooNature, ColonTobacco2017, Nandy2023Matter, ARCMOF, MajumdarDiversifyingACSAMI} 
These databases first began as exhaustive combinations of a small number of building blocks on a small number of topological nets\cite{Wilmer2012, BoydWooNature}, but later grew in diversity with increasingly complicated nets and building blocks.\cite{ColonTobacco2017,Nandy2023Matter, ARCMOF,MajumdarDiversifyingACSAMI} 
As MOF chemical space becomes increasingly sampled, constructing realistic MOFs\cite{MoosaviDiversity2020} that are synthetically accessible remains an open challenge. This challenge is further compounded by the fact that synthetically accessing a MOF is insufficient for evaluating its properties. 
Prior to its use as a functional material, solvent must be removed from the pores of the MOF. Additionally, the MOF must be thermally stable, stable to conditions with water, and mechanically stable to enable practical use.

\qquad With the rapidly expanding availability of experimental and synthetic data for MOFs, machine learning (ML) has emerged as a powerful, cost-effective, and efficient alternative to density functional theory (DFT) for designing functional MOFs.\cite{Fu2023,Lee2015,Nazarian2017,shengchao2023,Zhao2024PACMAN} 
Supervised ML models trained on high-throughput computational data have achieved remarkable success in predicting key properties such as adsorption capacities\cite{Altintas2021}, band gaps\cite{Rosen2022}, conductivity\cite{Dou2024}, and stability metrics such as mechanical stability.\cite{Moghadam2019,Nandy2023Matter} Additionally, ML models trained on mined experimental data have been able to predict activation stability\cite{NandyJACS2021, NandySciData2022}, thermal stability\cite{NandyJACS2021,NandySciData2022}, and most recently, water stability.\cite{Terrones2024}
Transformer-based models pretrained on extensive MOF datasets have further advanced the state-of-the-art in property prediction, significantly improving generalizability and data efficiency, particularly when large-scale datasets are available\cite{Cao2023,Kang2023MOFTransformer,Park2024}. 
Concurrently, generative models such as variational autoencoders\cite{Yao2021SMVAE}, generative adversarial networks\cite{Kim2020GAN}, and large language models\cite{Kang2024ChatMOF,Yaghi2023AugJACS,Yaghi2023JACS,YaghiReview2025} have enabled the \textit{de novo} design of MOFs tailored for specific functions. 
More recently, diffusion models\cite{ddpm}, which have significantly enhanced the quality of generated images\cite{iddpm}, molecules\cite{EDM}, and chemical reactions\cite{OAReactDiff,duan2024reactot}, have emerged as promising tools for directly sampling and generating three-dimensional MOF structures. 
For example, MOFDiff\cite{fx2023MOFDiff} employs a diffusion model to generate coarse-grained MOF structures—comprising the identities and central positions of building blocks—that are subsequently assembled using an algorithmic approach. 
This assembly step was later transformed into a learnable flow-matching process in MOFFlow\cite{kim2024mofflow}. 
However, both of these approaches primarily assemble existing building blocks without generating novel 3D structures for the building blocks themselves, which often constitutes the main mode of innovation from experimental chemists. 
Furthermore, the generation of materials using all-atom diffusion models typically faces size constraints of around 200 atoms, owing to computational limitations and GPU memory demands\cite{joshi2025ADiT}. Such a constraint has limited the all-atom generation of MOF unit cells, which frequently contain hundreds or thousands of atoms.
In a concurrent work\cite{inizan2025agenticMOF}, Inizan et al. developed the first agentic AI systems for end-to-end MOF generation and experimental validation, showing great promise for automated synthesizable MOF discovery. 

\qquad We address these challenges by introducing building-block aware (BBA) MOF Diffusion, a diffusion model framework that leverages a BBA representation of MOFs combined with SE(3) equivariance\cite{OAReactDiff,Yao2021SMVAE}.
Our BBA MOF Diffusion model was trained on the CoRE 2019 dataset, capturing the joint distribution of inorganic nodes, organic edges, and topological nets present in experimentally synthesized MOFs. 
Since our model directly learns from 3D all-atom representations of individual building blocks, it effectively generates novel inorganic nodes and organic edges absent in the training set. 
Additionally, by explicitly utilizing topological nets to represent repeating structural units, our approach significantly reduces the computational burden, allowing the generation of MOFs with exceptionally large unit cells up to approximately 1000 atoms. 
The MOFs produced by BBA MOF Diffusion exhibit high structural validity, substantial novelty, and uniqueness, while maintaining distributions comparable to those in established experimental databases. 
Finally, we experimentally synthesized $\mathrm{[Zn(1,4-TDC)(EtOH)_{2}]}$ as the target MOF, which had the expected topological structure, proving the reliability of our method.

%% file: 02_results.tex
\section*{2. Results and Discussions}

\begin{figure*}[t!]
    \centering
    \includegraphics[width=0.86\textwidth]{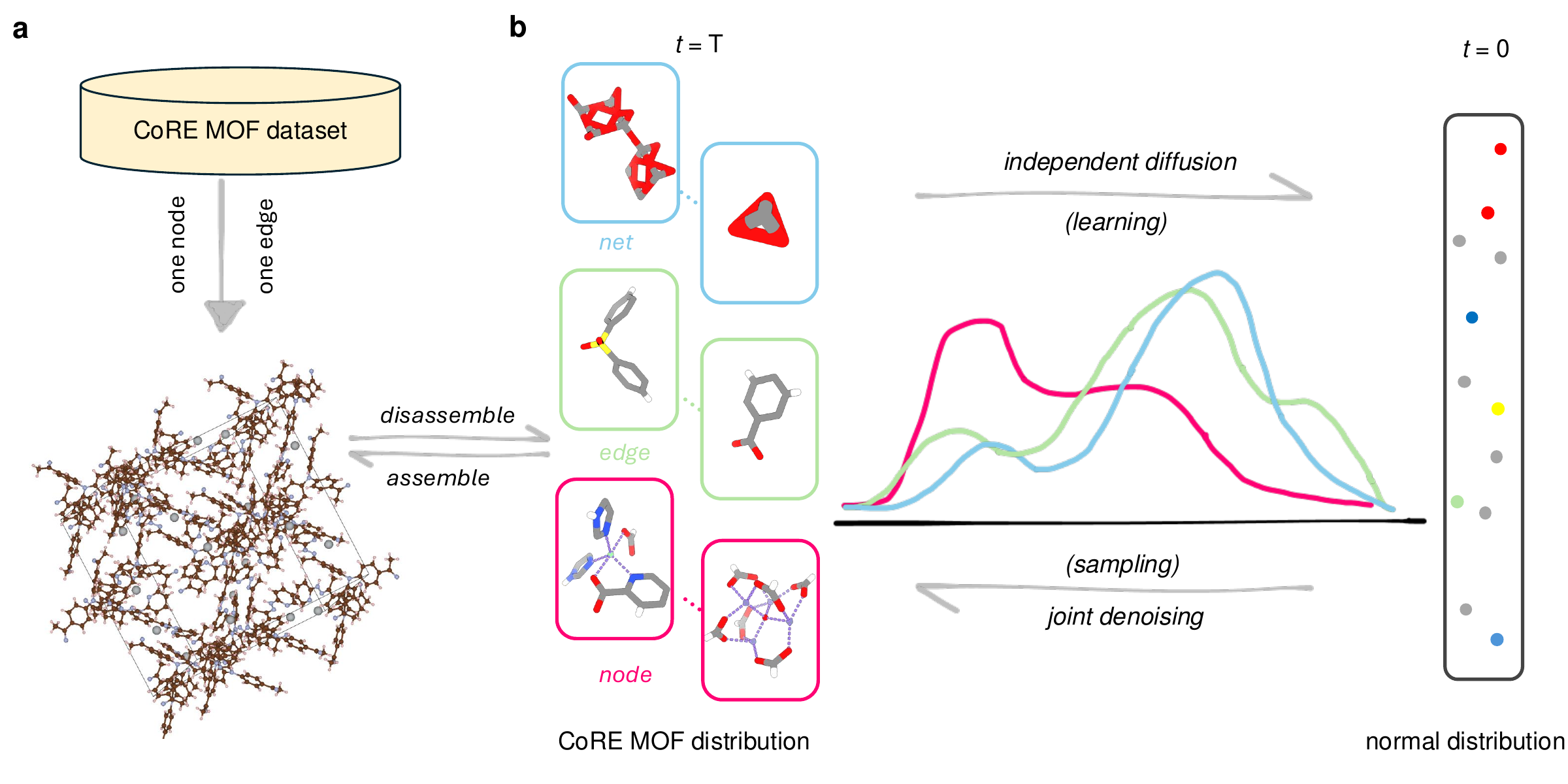}
    \caption{\textbf{Schematic for building block aware MOF diffusion models.}
    \textbf{a.} MOFs in the CoRE MOF 2019\cite{Chung2019} dataset composed of one inorganic node and one organic edge are selected and disassembled into their respective building blocks to be used for diffusion model training. During the sampling (generation) phase, the generated building blocks are assembled to form a MOF.
    \textbf{b.} Learning the joint distribution of three building block components in a MOF: 1) inorganic node (red), 2) organic edge (green), and 3) topological net (blue). A forward independent diffusion process corrupts the joint distribution of CoRE MOF dataset at $t=T$ to an independent normal distribution at $t=0$, during which an object-aware SE(3) GNN is trained with the denoising objective function. In the backwards direction, this scoring network is applied to denoise from samples in the normal distribution to the original joint distribution.
    In inorganic nodes and organic edges, atoms are colored as follows: Zn in purple, Cu in green, C in gray, N in blue, O in red, S in yellow, and H in white.
    Gray for inorganic node and red for organic edge for coloring the topological nets.
    }
    \label{fig:overview}
    
\end{figure*}

\paragraph{2a. Overview of BBA MOF Diffusion.}
Using a denoising score matching objective, diffusion models learn the underlying analytically non-tractable distribution of observed samples by training a scoring network (see \textit{\nameref{bba_mof_diffusion}}); \cite{diffusion2015, ddpm, scoresde}). 
Duan et al. \cite{OAReactDiff} further extended these models to accommodate chemical systems composed of multiple three-dimensional objects with non-spatial interactions by employing object-aware SE(3)-equivariant graph neural networks, which accurately enforce all necessary symmetries. 
Similarly, metal–organic frameworks (MOFs) can be transformed from their unit cell representations into building block representations expressed in the language of nodes, edges, and topological nets\cite{pormake, Bucior2019}, making MOFs an ideal playground for object-aware diffusion models.\cite{Nandy2023Matter, Moghadam2019}

\qquad Following prior work\cite{Nandy2023Matter}, MOFs in the CoRE MOF 2019\cite{Chung2019} database were deconstructed into their constituent building blocks, using an automated deconstruction algorithm to separate inorganic nodes, organic nodes, and organic edges.\cite{Nandy2023Matter} This algorithm depends on the interpreted connectivity matrix based on pairwise atomic distances in a given .CIF file.\cite{MoosaviDiversity2020} MOFs with interpenetration (i.e. two interweaved but separable unit cells) are not considered in this work. A node is defined as a group of atoms that have more than two connection points. Although MOFs can have both inorganic and organic nodes, in this work we focus only on MOFs containing inorganic nodes. Edges are building blocks with two connection points that connect nodes on a given topological net. To identify the set of building blocks used in the training data, we curate a list of nets with one inorganic node and one edge (964 total nets), and identify all MOFs that readily fit on at least one of these topological nets, using heuristic criteria.\cite{Nandy2023Matter,pormake}. Any inorganic nodes that met a strict RMSD criterion of < 0.3\AA threshold on a given single inorganic node-single organic edge net were used as training data. Additionally, any edges with two connection points and angles $\geq$ 120\degree were used for generation. During building block generation and MOF construction, we restrict ourselves to 4 simple topological nets that are a subset of nets studied in prior work\cite{Wilmer2012}: \textit{dia}, \textit{nbo}, \textit{sra}, and \textit{pcu}. 
This restriction on simple nets is in part motivated by the substantial presence of simple topological nets in experimentally synthesized MOFs.\cite{YaghiChemRevNets2014, EddaoudiChemRevNets2020} 
Although the focus of hypothetical databases is frequently to increase the diversity of sampled chemistry, hence venturing into novel topologies, our focus is on generating MOFs that are from the distribution of experimentally synthesized materials.

\qquad Using this curated dataset that represents the distribution of building blocks in experimentally synthesized MOFs, we trained a BBA MOF Diffusion model using LEFTNet\cite{leftnet}—a state-of-the-art equivariant graph neural network—as the scoring network. 
The training process involves adding Gaussian white noise to the inorganic nodes, organic edges, and topological nets from 
$t=0$ to $t=T$, enabling the model to learn the joint distribution of CoRE MOFs (Fig. \ref{fig:overview}b).
Once trained, the model can sample nodes, edges, and topological nets from a normal distribution; these samples are then propagated backward through the denoising process using the learned scoring network (i.e., LEFTNet), thereby recovering building blocks that closely resemble those in the CoRE MOF database. 
The recovered components are subsequently assembled using PORMAKE \cite{pormake} to construct the three-dimensional structure of a MOF.
This workflow confers several advantages over previous generative models for MOFs. 
First, by utilizing the building block representation, the effective number of atoms the diffusion model must generate is drastically reduced compared to the direct generation of an explicit unit cell, resulting in a higher success rate.
Second, when trained on experimentally synthesized MOFs, the BBA MOF Diffusion model can generate novel building blocks (e.g., edges), thereby transcending the limitations imposed by existing components. 
Lastly, the model inherently preserves key conceptual elements—such as nodes, edges, and nets—that experimentalists routinely employ in MOF design.
Combined with a flexible conditional generation workflow which allows partial structure design and composition, these advantages make the BBA MOF Diffusion model a practical and powerful tool for chemists seeking to design novel MOFs.

\begin{figure*}[t!]
    \centering
    \includegraphics[width=0.98\textwidth]{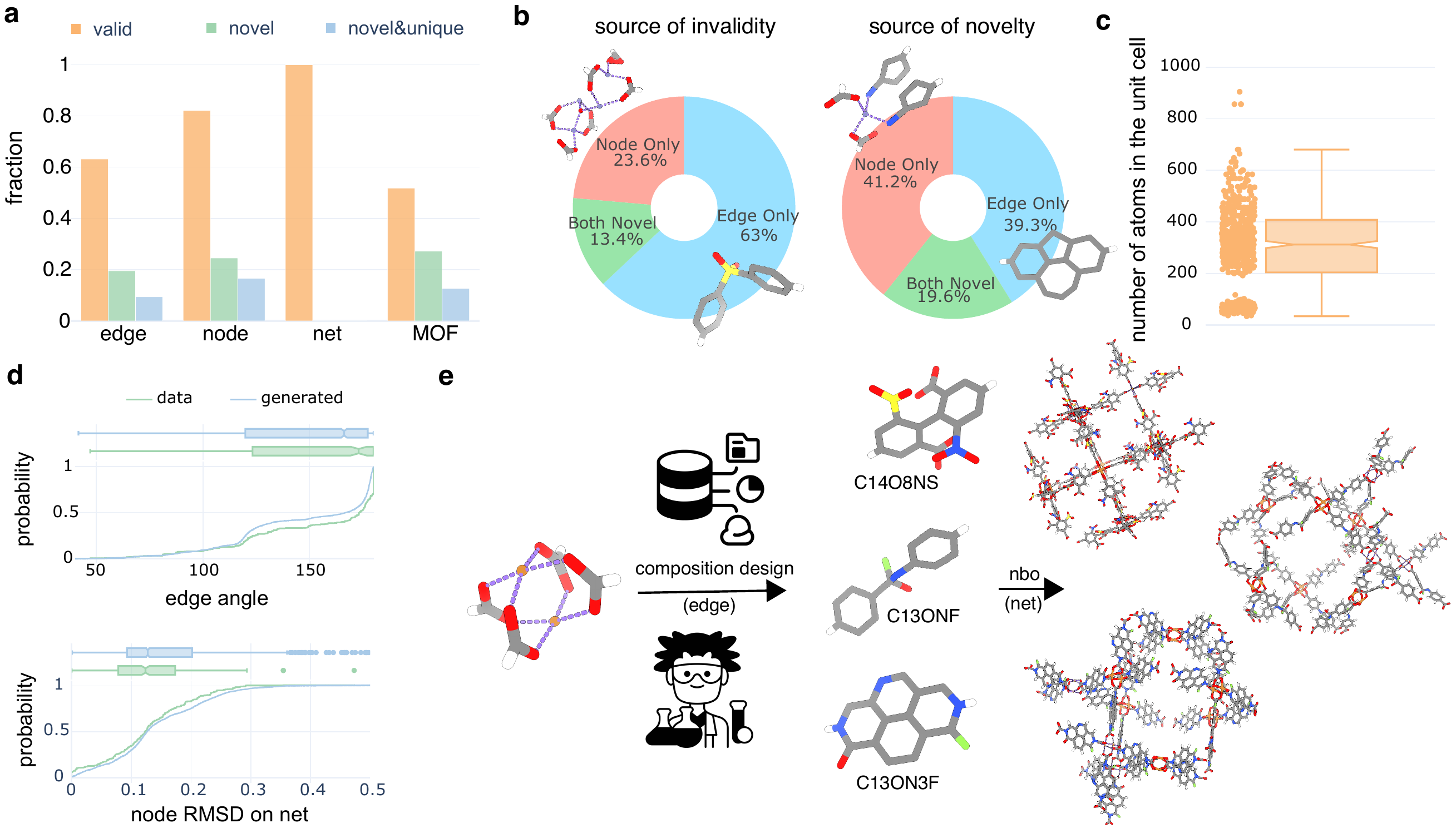}
    \caption{\textbf{Statistics for the generated MOFs.}
    \textbf{a.} Validity (orange), novelty (green) and combined (blue) for the three building blocks (edge, node, and net) independently and the final MOF assembled.
    \textbf{b.} Source of invalidity (left) and novelty (right) broken down to the contribution of node, edge, or both.
    \textbf{c.} Distribution for the number of atoms in the unit cell of generated MOFs. Both all scatters and a notched box plot are shown.The quarter 1 and 3 are the edges of the box, and fences corresponding to the edges +/- 1.5 times the interquartile range.
    \textbf{d.} Cumulative probability for the edge angle (top) and node RMSD after net optimizations (bottom) for the CoRE MOF dataset (green) and generated MOFs (blue). A box plot for the same distribution is shown at the top margin, correspondingly.
    \textbf{e.} Conditional MOF generation with dicopper tetracarboxylate “paddle-wheel” cluster as the known inorganic node. Desired chemical compositions for edge design can be proposed, with examples as C14O8NS (top), C13ONF (middle), and C13ON3F (bottom). With edge composition as inputs, BBA MOF diffusion conditionally samples 3D structures for organic edges, which are assembled together with the known inorganic node and topological net (here, nbo) to the final MOF structures.
    }
    \label{fig:stats}
\end{figure*}

\begin{figure*}[t!]
    \centering
    \includegraphics[width=0.98\textwidth]{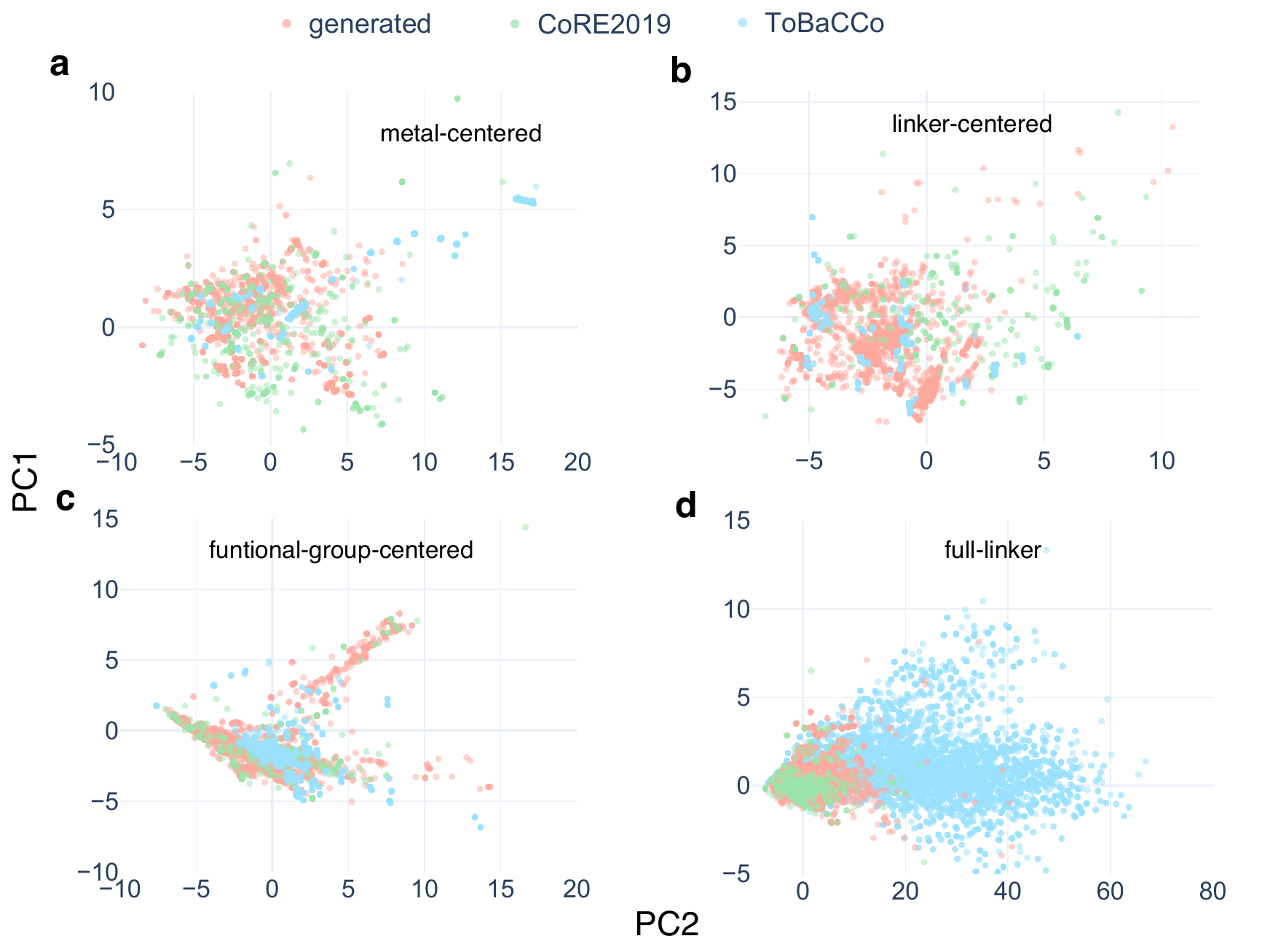}
    \caption{\textbf{Overlaid revised autocorrelation (RAC) features of MOFs on RAC subsets.} Opacity is utilized to visualize data overlap, and z-order is selected for visualization.
    \textbf{a.} Metal-centered RAC features for BBA Diffusion generated MOFs (red), CoRE MOF 2019 (green), and ToBaCCo MOF (blue). 
    \textbf{b.} Linker-centered RAC features. 
    \textbf{c.} Non-C functional-group-centered RAC features.
    \textbf{d.} Full-linker RAC features.
    }
    \label{fig:RACcomparison}
\end{figure*}

\paragraph{2b. Generating Novel MOFs.}
We apply the BBA MOF Diffusion model to generate MOFs using a sampling scheme referred as \textsc{separate} mode, where each of the three building blocks is independently sampled from the building block library.
This mode delivers a balance treatment between diversity and synthesizablity of generated MOFs.
It is noted, however, there is no guarantee that the resulting combination corresponds to any existing MOF, although individual building blocks may be present in the database.
For each combination of building blocks, eight samples were generated for roughly 1200 sets of randomly selected compositions, resulting 9,712 MOFs in total.
A relatively high validity of generated edges (63\%) and nodes (82\%) is observed in the MOFs sampled, demonstrating the robustness of BBA MOF diffusion model for generating building blocks (Fig. \ref{fig:stats}a).
As only four distinct topological nets are included in the training data, the model simply repeats the four nets and thus with 100\% validity. 
This behavior is desired, since the majority of synthesis efforts target existing MOF topologies rather than designing new ones.
A MOF is categorized as valid only if all building blocks are valid, resulting 52\% rate of overall validity for generated MOFs.
Looking into the source of invalidity, it is observed that failure in edge generation is twice as frequent as that in node generation (Fig. \ref{fig:stats}b).
This is mostly caused by a tight constraint on the connecting points angle between 140 and 180 degree, which is inherited from a previous work\cite{Nandy2023Matter}, mimicking an ideal linear arrangement for the two connecting points in an edge (Supplementary Figure \ref{Supp:edge_source_invalidty}).
In reality, however, an edge could still form a valid MOF in spite of its acute angle, with a salient example being sulfonyldibenzene, which frequently appears in CoRE MOF database but has an angle between connecting points that is < 140 degree.
A significant fraction of invalid inorganic nodes originates from unphysical 3D geometries with floating atoms or unphysical bonding patterns, showcasing the difficulty of generating inorganic nodes in MOFs, which usually have more complex chemistry  (Supplementary Figure \ref{Supp:node_source_invalidty}).
Out of these valid samples, 20\% of the edges and 25\% of the nodes are novel (i.e., having a different connectivity), which shows that the model is more than memorizing the building blocks that it has seen in the training data (Fig. \ref{fig:stats}a).

\qquad A MOF is considered novel when there is at least one novel building block, resulting an overall novelty rate of 27\%, slightly higher than that of both edges and nodes.
There, an equal contribution of source of novelty for edges and nodes is observed, which suggests a balanced treatment of both building blocks in our model (Fig. \ref{fig:stats}b).
More importantly, out of 27\% novel MOFs, only less than half are duplicates, resulting 13\% MOFs that are both novel and unique, making them potential candidates for experiment validation.
Interestingly, highly similar distributions of edge angles and root mean square deviation (RMSD) of nodes after MOF assembly are observed, suggesting the model learns the underlying characters of the CoRE MOF data (Fig. \ref{fig:stats}d).
With BBA MOF diffusion, an extremely wide range of number of atoms in the unit cell, ranging from 37 to 904, is found in generated MOFs.
This is in huge contrast compared to methods that attempt to generate the entire unit cell as a whole, which is currently limited to relatively small unit cells with dozens of atoms, due to the algorithmic complexity and computational cost of generating large unit cells with symmetry (Fig. \ref{fig:stats}c).
It is worth noting that the size of the generated MOF naturally becomes larger with more complicated topological nets and is irrelevant to the actual graph that is internally handled by the scoring network, ensuring the scalability of BBA MOF diffusion to extremely large and complex MOF structures (Supplementary Figure \ref{Supp:natoms_mof_vs_graph}).

\qquad Experimentalists commonly design MOFs with the concepts of building blocks by either top-down or bottom-up approach.
Frequently, new edges are desired with previously known metal nodes and topological nets to achieve new functionality or improve certain properties.
BBA MOF diffusion model as an approach that adopts build block representation for MOFs can readily be applied for this conditional generation task with the inpainting technique (see \textit{\nameref{bba_mof_diffusion}}).
Here, we use dicopper tetracarboxylate “paddle-wheel” cluster, which commonly present with the \textit{nbo} net, as an example for conditional generation of MOFs (Fig. \ref{fig:stats}e).
With the node and net fixed, one would further attempts multiple chemical compositions of the organic edges, by either searching public databases (e.g., PubChem) or designing fully by the need, as some physical and chemical properties are relatively correlated with compositions for organic chemistry.
For example, with a chemical composition of C$_{13}$ONF, a fluorinated anilino(phenyl)methanol edge is generated for the combined MOF.
Although the fluorinated variant of anilino(phenyl)methanol is absent in PubChem, the sp$^3$ carbon is indeed a position that is more likely to be functionalized in anilino(phenyl)methanol.
One can also be creative and propose chemical compositions that are rare in open-source databases, such as those with heavy atom backbone of C$_{14}$O$_8$NS. 
The BBA MOF diffusion model still generates a reasonable organic edges where a MOF structure can be successfully assembled with a few attempts.
In contrast, for highly abundant composition in PubChem such as C$_{13}$ON$_3$F, the model would repeatedly generates edge structures that are most likely to form MOFs with the dicopper tetracarboxylate “paddle-wheel” cluster, regardless of a huge potential design space for edges.

\paragraph{2c. Diversity Distribution of Generated MOFs}

\qquad To analyze the distributions of MOFs constructed using our BBA MOF diffusion algorithm, we quantify the diversity of subsets of MOF revised autocorrelation (RAC) features, following our prior work.\cite{MoosaviDiversity2020} 
RACs are pairwise products and differences on the atom-attributed molecular graph that fingerprint the chemical environment from various scopes and localities.\cite{JanetJPCA2017} 
The resulting high-dimensional graph-based fingerprints are projected into two dimensions using principal component analysis (PCA) for the simplicity of visualization. 
Generated MOFs with BBA MOF diffusion faithfully represent MOFs in the CoRE MOF database and closely mimic the distributions of metals, linkers, and coordination environments present in the CoRE MOF database(Fig. \ref{fig:RACcomparison}). 
Topologically diverse databases such as ToBaCCo MOF\cite{ColonTobacco2017}, on the other hand, demonstrate increased diversity for the linker connectivity (i.e. linkers with more complex connectivity to fit more complex topological nets) at the cost of deviating from the distribution of the CoRE MOF database (Fig. \ref{fig:RACcomparison}d).
In addition, the metal and coordination environments in ToBaCCo dataset are largely constrained compared to CoRE MOFs, partially due to the difficulty of proposing hypothetical metal nodes (Fig. \ref{fig:RACcomparison}a). Correspondingly, hypothetical databases such as ToBaCCo frequently select a few commonly studied metal nodes.\cite{ColonTobacco2017,MoosaviDiversity2020}
In comparison, MOFs generated by BBA MOF diffusion models may be more suitable to discover new, unexplored MOFs that are characteristically distinct from existing structures but still synthetically accessible as they resemble a more similar distribution to historically synthesized MOFs. 

\begin{figure*}[t!]
    \centering
    \includegraphics[width=0.98\textwidth]{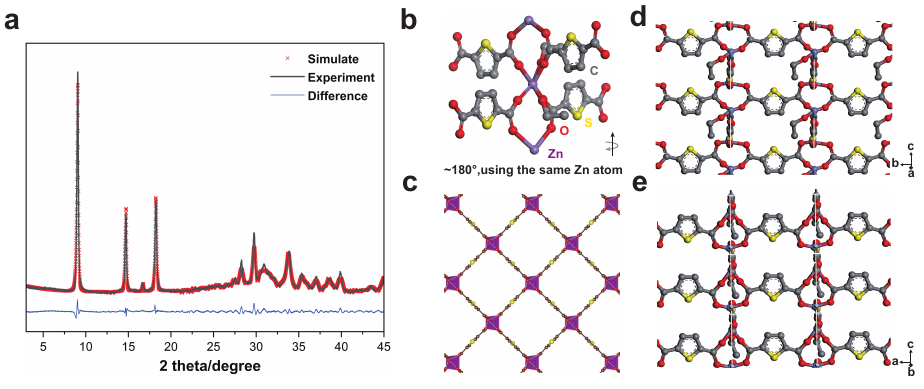}
    \caption{\textbf{PXRD, structural refinement results and the detailed structure of $\mathrm{[Zn(1,4-TDC)(EtOH)_{2}]}$ } 
    \textbf{a.} Experimental PXRD pattern of $\mathrm{[Zn(1,4-TDC)(EtOH)_{2}]}$ and its structural refinement results, The red cross, black line and blue line represent the refinement results, experimental results and differences, respectively.
    \textbf{b.} Diagram of the SBU structure of $\mathrm{[Zn(1,4-TDC)(EtOH)_{2}]}$. 
    \textbf{c.} 
    \textbf{d.} 
    \textbf{e.} 
     Views of the$\mathrm{[Zn(1,4-TDC)(EtOH)_{2}]}$ MOF crystal structure from the C-axis, A-axis, and B-axis directions, respectively.
    }
    \label{fig:4}
\end{figure*}

\paragraph{2d. Synthesis of new MOFs.}
\qquad 
To verify the feasibility of BBA MOF diffusion-guided MOFs, we synthesized a MOF corresponding to its suggestions for metal nodes, organic ligands, and network structures. 
A high scoring structure $\mathrm{[Zn(1,4-TDC)(EtOH)_{2}]}$ predicted by the model was synthesized using the common solvothermal synthesis method. Meanwhile, because of the participation of small solvent molecules in the coordination of the MOF structure, the final structure of the MOF shows a strong solvent-dependent characteristic. \cite{2013Influential,2007Tri}
We carried out the synthesis in the ethanol system at 130 °C and obtained the MOF structures of special SBU and net structure.
We completed the PXRD test and constructed the target structure of the MIL-53 topology. 
The refinement result shows that the simulation results fit well with the powder diffraction results of the MOF, ensuring the authenticity of the simulated structure (Fig. \ref{fig:4}a). 
From the point of view of SBU, the new MOF structure is composed of Zn(II) nodes bridged by 2,5-thiophenedicarboxylate  (TDC) ligands. 
The central Zn is pseudo-octahedrally coordinated by TDC and ethanol sitting on a crystallographic center of inversion. 
There are four TDC ligands coordinated to each trimetal node. The tow carboxylates each bridge the central atom to the two terminal Zn atoms, which are in pseudo-tetrahedral environments. 
The apical position is occupied by ethanol (Fig. \ref{fig:4}b)/
An hexagonal node then results in the formation of a 2-D triangular network. 
Stacking of adjacent layers proceeds in an ABAB fashion in the (001) direction (Fig. \ref{fig:4}c-e).

\qquad We obtained the optimal growth conditions by adjusting the concentration of reactants and the reaction temperature to get rod like crystals (Supplementary Figure \ref{Supp:sem}).
TGA of the $\mathrm{[Zn(1,4-TDC)(EtOH)_{2}]}$ shows the quality changes of the two divisions. 
The first part of quality loss occurs between 130 °C and 160 °C resulting from the loss of ethanol and water from the network void space. 
A second weight loss of 40 centered at 400 °C corresponds to network decomposition. The remaining 52\% wt., which is stable to 700 °C is attributed to formation of ZnO (Supplementary Figure \ref{Supp:tga}).
The {$\mathrm{N_2}$} Adsorption-desorption Analysis was also experiment to obtained the BET surface areas of 140.37 ± 4.80 m²/g and a pore size of approximately 1.4 nm (Supplementary Figure \ref{Supp:bet1} -\ref{Supp:bet2}).
Interestingly, the experimentally confirmed structure differs slightly to the model suggested one, where two TDC ligands that were suggested to coordinate with Zn are actually replaced by ethanol.
This change can be understood by the fact that all solvents are removed from the MOF structure during the curation of the original CoRE 2019 database, leading to lost of information where no model can learn.
Still, a new MOF of similar structure was synthesized inspired by MOFs generated by BBA MOF diffusion, indicating the great promise from generative AI-guided materials design.

%% file: 03_discussion.tex
\section*{3. Conclusions and Outlook}
Designing novel building blocks and their topological connectivity is crucial for discovering stable metal-organic frameworks (MOFs) with diverse functionalities and applications. 
In this work, we introduce BBA MOF diffusion, a diffusion model trained to capture the joint distribution of building blocks, enabling the generative design of three-dimensional MOF crystal structures. 
Importantly, BBA MOF diffusion can propose previously unseen nodes and edges, serving as a fundamental source of creativity in MOF design.
By employing a building-block-based representation, MOFs are systematically assembled through nodes and edges consistent with the given topology, enabling scalability in the generated structures. 
The experimental synthesis and validation of a MOF exhibiting the targeted $\mathrm{[Zn(1,4-TDC)(EtOH)_{2}]}$ property demonstrate the practicality and potential of BBA MOF diffusion in facilitating effective human-AI collaboration.

\qquad Nevertheless, several limitations of the current approach must be recognized. 
The present model is restricted to MOFs containing a single inorganic node and one organic edge, thus constraining the variety of structures it can explore and generate. 
While extending the model to incorporate additional nodes and edges is feasible, this expansion will inevitably confront the same size limitation (ca 200 atoms per graph) inherent to all-atom diffusion models\cite{joshi2025ADiT,inizan2025agenticMOF}.
Additionally, due to the long-tail distribution of topological nets in MOF databases, our training set was limited to the four most prevalent nets. 
Consequently, although our model generates novel chemical configurations for nodes and edges, it does not currently learn to generate entirely new topological nets. 
A more comprehensive and diverse dataset encompassing a broader range of topological nets would greatly enhance the model's ability to derive generalized design principles for topology in MOFs.
Moreover, while structural conditional generation via inpainting has been implemented, property-driven conditional generation remains unaddressed in the current iteration of BBA MOF diffusion. 
Such capability would be vital for tailored MOF design targeting specific applications. Future extensions of our model could incorporate classifier-free, property-guided diffusion strategies\cite{clf_free_diffusion}, a development we plan to pursue in a future work.
Lastly, chemical compositions are assumed known prior to the use of BBA MOF Diffusion to incorporate intuition of chemists, which may not hold in complete free generation scenario.
There, a composition generation model (for example, LLM-based LinkerGen in ref.\cite{inizan2025agenticMOF}) can be adapted before applying BBA MOF Diffusion for the generation of 3D structures.

\qquad BBA MOF diffusion underscores that non-spatial interactions—such as the statistical likelihood of combinations of nodes, edges, and topological nets forming stable MOFs—can indeed be effectively encoded and learned within an object-aware SE(3)-equivariant diffusion framework. 
Beyond MOFs and chemical reactions\cite{OAReactDiff}, many other systems in physical sciences, including co-crystals, polymeric assemblies, protein-ligand complexes, antibody-drug conjugates, and others characterized by non-spatial, causal relationships among distinct components, could similarly benefit from this building-block-aware diffusion approach. 
Carefully tailored implementations of the method introduced here thus hold significant potential for broad and insightful applications across diverse scientific domains. 

%% file: 04_methods.tex
\section*{4. Materials and Methods}

\paragraph{4a. Details for BBA MOF Diffusion}\label{bba_mof_diffusion}
\textit{Symmetry in modeling MOFs.-- }
A function $f$ is said to be equivariant to a group of actions $G$ if $g\circ f(x) = f(g\circ x)$ for any $g\in G$ acting on $x$~\cite{equivariance, gdlbook}.
We specifically consider the Special Euclidean group in 3D space (SE(3)) which includes translation and rotation transformations with additional permutation transformations, which are symmetries that any 3D objects should follow.
We intentionally break reflection symmetry so that our model can describe molecules with chirality.
For systems that consist of multiple objects (e.g., three building blocks in MOFs) that do not have direct interactions in the 3D Euclidean space, we adopt the object-aware SE(3) symmetry established in our previous work\cite{OAReactDiff} to fulfill the symmetry required while maintaining expressiveness to describe the target systems.

\textit{Equivariant diffusion models.-- }
Diffusion models are originally inspired from non-equilibrium thermodynamics~\cite{ddpm, diffusion2015,scoresde}.
A denoising diffusion probabilistic model (DDPM) has two processes, the forward (diffusing) process and the reverse (denoising) process.
In the forward direction, the noising process gradually adds white noise into the data until it becomes a prior (Gaussian) distribution that is more readily sampled:
$$
q(x_t|x_{t-1}) = \mathcal{N}(x_t|\alpha_t x_{t-1}, \sigma_t^2 I),
$$
where $\alpha_t$ controls the signal retained and $\sigma_t$ controls the noise added.
A signal-to-noise ratio is defined as $\text{SNR}(t)=\frac{\alpha_t^2}{\sigma_t^2}$. We set $\alpha_t = \sqrt{1-\sigma_t^2}$ following the variance preserving process in~\cite{scoresde}.

\qquad The \textit{true denoising process} can be written in a closed form due to the property of Gaussian white noise:
\begin{align}
q(x_s|x_0, x_{t}) &= \mathcal{N}(x_s|\mu_{t\rightarrow s}(x_0, x_t), \sigma^2_{t\rightarrow s}I), \nonumber\\
\mu_{t\rightarrow s}(x_0, x_t) &= \frac{\alpha_{t|s}\sigma^2_s}{\sigma^2_t}x_t + \frac{\alpha_s\sigma^2_{t|s}}{\sigma^2_t}x \;\; \text{and} \;\; \sigma_{t\rightarrow s} = \frac{\sigma_{t|s}\sigma_{s}}{\sigma_t}, \nonumber
\end{align}
where
\textcolor{black}{
$s$ < $t$ refers to two different timesteps along the diffusion/denoising process ranging from 0 to T,
}
$\alpha_{t|s} = \frac{\alpha_{t}}{\alpha_{s}}$, $\sigma^2_{t|s} = \sigma^2_t - \alpha^2_{t|s} \sigma^2_{s}$.
However, this \textit{true denoising process} is dependent on $x_0$, which is the non-accessible data distribution.
Therefore, diffusion learns the denoising process by replacing $x_0$ with $x_t + \sigma_t \epsilon_{\theta}(x_t, t)$ predicted by a denoising network $\epsilon_{\theta}$, which predicts the difference of $x$ between two time steps (that is, $\epsilon$).
The training objective is to maximize the variational lower bound (VLB) on the likelihood of the training data:
$$
-\log p(x) \leq D_{KL}(q(x_T|x_0)||p_{\theta}(x_T)) - \log p(x_0|x_1) + \Sigma_{t=2}^T D_{KL}(q(x_{t-1}|x_0, x_t)|| p_{\theta}(x_{t-1}|x_t))
$$
Empirically, a simplified objective function has been found to be efficient to optimize~\cite{ddpm}:
$$
\mathcal{L}_\mathrm{simple} = \frac{1}{2}||\epsilon - \epsilon_{\theta}(x_t, t)||^2,
$$

\textit{LEFTNet.-- }
We build our object-aware SE(3) transition kernel on top of a recently proposed SE(3)-equivariant GNN, LEFTNet~\cite{leftnet}. 
LEFTNet achieves SE(3)-equivariance by building local node- and edge-wise equivariant frames that scalarize vector (e.g. position, velocity) and higher order tensor (e.g. stress) geometric quantities. 
These geometric quantities are transformed back from scalars through a tensorization technique\cite{leftnet} without loss of any information. LEFTNet is designed to handle Euclidean group symmetries including rotation, translation and reflection, as well as the permutation symmetry. 
We adopt the object-aware improvement from our previous work, OA-ReactDiff\cite{OAReactDiff}, to tailor this model for MOFs. 
For more detailed descriptions, we refer reading the original LEFTNet\cite{leftnet} or OA-ReactDiff\cite{OAReactDiff} works.

\textit{Model training.-- }
During BBA MOF diffusion, each building block is represented by atom types with one-hot encoding. 
The constituent atoms are represented as nuclear charges and Cartesian coordinates.
With the assumption that chemists would prefer to provide chemical compositions for MOFs under design, we only diffuse and denoise the Cartesian coordinates of each building block while keeping the atom types constant.
It should be noted, however, this is a choice rather than limit of the diffusion model framework, as diffusing and denoising atom types have been successfully demonstrated using discrete diffusion in other work \cite{campbell2022continuous}.
We trained the BBA MOF diffusion model with LEFTNet as our transition kernel.
We used the same set of hyperparameters to that for our previous work of OA-ReactDiff\cite{OAReactDiff}, with 96 radial basis functions, 196 hidden channels for message passing, and 6 equivariant update blocks. 
A large neighbor cutoff threshold of 10\AA\, is used to enhance expressiveness for large building blocks.
We also adopted hyperparameters of the diffusion process from the OA-ReactDiff paper, where a \textcolor{black}{second-order} polynomial noise schedule \textcolor{black}{(that is, $\alpha_t = 1 - (t / T)^2$)} and $L_{\mathrm{simple}}$ loss function is used with a total timestep of 5000.
We used a learning rate of 0.0005 and a batch size of 16, which is the largest batch size that we can afford with a A30/24GB GPU for training 2000 epochs.
All data was used for training without any train/test partitioning as there are no in-distribution metrics examined.

\textit{Conditional generation.}
Inpainting is a flexible technique to formulate the conditional generation problem for diffusion models. \cite{Repaint}
Instead of modeling the conditional distribution, inpainting models the joint distribution during training.
During inference, inpainting methods combine the known and diffused part as the context through the noising process of the diffusion model before denoising the combined input.
The resampling technique~\cite{Repaint} has demonstrated excellent empirical performance in harmonizing the context of the denoising process as there is sometimes mismatch between the noised conditional input and the denoised inpainting region. 
One distinct character of inpainting is that it is guaranteed to recover the known part at the end of the sampling process, in favor of MOF design processes where certain building blocks are rigidly required.

\paragraph{4b. Dataset curation.}\label{data_curation}
The CoRE MOF 2019 database is a sanitized database of experimental structures. We construct our building-block dataset from this database, using code from prior work.\cite{Nandy2023Matter} To deconstruct a MOF, we first interpret the adjacency matrix using pairwise distance cutoffs.\cite{MoosaviDiversity2020} After this, we separate the MOF into inorganic nodes, organic nodes, and edges, using an automated deconstruction algorithm. Nodes, both organic and inorganic, are defined as building blocks with three or more connection points. These connection points may be attached to other nodes or organic edges. Edges are defined as organic molecules with two connection points.

\qquad In this work, we narrow our focus to MOFs that have a single inorganic node and a single edge. These are the simplest types of MOFs. Therefore, we first isolate the set of MOFs that have one distinct inorganic node and one distinct organic edge, although the stoichiometry between these two may vary. Distinctness is quantified by identifying the Weisfeiler-Lehman graph hash on the atom-wise attributed adjacency matrix.  

\qquad After identifying this subset of MOFs, we identify compatibility with single-node-single-edge MOF topological nets, since the CoRE MOF 2019 database does not store topological net information, and net assignment requires analysis with software such as TOPOS Pro which are not amenable to high throughput labeling. Using a cutoff of 0.3 \AA, as in prior work,\cite{Nandy2023Matter, LeeACSAMI2021} we determined compatibility of inorganic node building blocks with topological nets. We initially determined organic building blocks using a 140\degree cutoff angle.
For all building blocks, we omit hydrogen atoms as many of them are not characterized with resolution required as well as to reduce the size of MOFs to be handled by diffusion models.

\paragraph{4c. Details for experiments}
\textit{Chemicals.-- }
Ethanol was purchased from Aladdin(Shanghai, China). 2,5-thiophenedicarboxylic acid and zinc acetate dihydrate were purchased from Bidepharm(Shanghai, China). Unless otherwise specified, all chemicals do not require further purification before use.

\textit{{$\mathrm{N_2}$} Adsorption-desorption Analysis.-- }
The nitrogen adsorption measurements were performed on a Micromeritics ASAP 2020 plus Surface 
Area and Porosity Analyzer. As determined by the thermogravimetric analysis, the sample was heated to 90 °C, 
under a dynamic vacuum of 4 mTorr until the outgas rate was less 
than 2 mTorr/minute. After this degas procedure. The evacuated sample tube was weighed again and the sample mass was 
determined by subtracting the mass of the previously tared tube. Using a liquid nitrogen bath (77 K), the {$\mathrm{N_2}$} isotherm was then measured. Ultra-high purity grade (99. 999\% purity) {$\mathrm{N_2}$}, oil-free valves
and gas regulators were used for all free space correction and measurement. 
Fits to the Brunauer-Emmett-Teller (BET) equation satisfied the published consistency criteria.

\textit{Powder x-ray Diffraction (PXRD).-- }
Rigaku Miniflex600 ({$\mathrm{\lambda}$} = 1.5406  Å) was used to collect the PXRD patterns of the samples during screening the reaction conditions. 

\textit{Scanning Electron Microscopy (SEM).-- }
SEM was conducted on Phenom XL G2 SEM and an operating 
voltage of 4 kV.

\textit{Thermogravimetric analysis (TGA).-- }
Thermogravimetric analysis (TGA)  was carried out on a TGA-Q600  in nitrogen atmosphere using a 10 °C/min ramp without equilibration delay. 

\textit{Synthesis of {$\mathrm{[Zn(1,4-TDC)(EtOH)_{2}]}$}.-- }
An ethanolic solution (15mL) of $\mathrm{[Zn(CH_{3}COO)_{2}\cdot 2H_{2}O]}$ (1.0 equiv, 219.51 mg) and $\mathrm{1,4-TDCH_{2}}$  (1.0 equiv, 172.16 mg) was placed in a teflon sleeve of a parr bomb. The parr bomb was then closed and left at 130 ° C for 24 hours.  After cooling to room temperature, the bomb was opened. The resulting white powder was centrifuged and washed twice with water, ethanol, and acetone.

%% file: 05_others.tex
\section*{Supporting Information} 
Distribution for sources of invalidity in generated edges and nodes;
Number of atoms for generated MOFs by size and graph;
SEM micrographs of [Zn(1, 4 - TDC)(EtOH)2]; 
TGA plot for [Zn(1, 4 - TDC)(EtOH)2];
Nitrogen adsorption isotherms for [Zn(1, 4 - TDC)(EtOH)2] at 77K; 
Barrett-Joyner-Halenda (BJH) pore size distribution for [Zn(1, 4 - TDC)(EtOH)2]

\section*{Code and data availability}
Code for training BBA MOF diffusion is open sourced and available at https://github.com/chenruduan/OAReactDiff. 
Dataset used for training BBA MOF diffusion model is available at https://zenodo.org/records/15308950.

\section*{Author contributions}
C.D., A.N.: conceptualization, methodology, software, validation, investigation, data curation, writing of original draft, review and editing, and visualization.
S.L.: experiment, investigation, writing of original draft, review and editing, and visualization.
Y.D.: methodology, software, and review and editing.
L.H.: review and editing.
Y.Q.: conceptualization, review and editing.
H.J.: review and editing. 
J.D.: conceptualization, methodology, validation, investigation, writing of original draft, review and editing.

\section*{Acknowledgement}
C.D. and H.J. would like to thank our entire team from Deep Principle for helpful discussions and support. 
A.N. gratefully acknowledges support from the Eric and Wendy Schmidt AI in Science Postdoctoral Fellowship, a Schmidt Sciences, LLC program.
C.D. and Y.Q. would like to thank M. Monkey for support.
S.L and J.D. gratefully thanks the National Key R\&D Program of China (Grant No. 2023YFE0206400), the National Natural Science Foundation of China (Grant No.22171185,No.12488201), the Fundamental Research Funds for the Central Universities, the staff of beamlines BL17B1 and BL14B1 of the National Facility for Protein Science Shanghai (NFPS) at the Shanghai Synchrotron Radiation Facility (SSRF).

\section*{Competing interests}
A.N. was formerly a paid consultant for FL96 by Flagship Pioneering, a venture focused on commercializing AI/ML-driven materials design.

%% file: 09_appendix.tex
\appendix

\title{\textit{Supplementary Information} for "Building-Block Aware Generative Modeling for 3D Crystals of
Metal Organic Frameworks"}

\maketitle

\renewcommand{\thesection}{\arabic{section}}  
\renewcommand{\thetable}{\arabic{table}}  
\renewcommand{\thefigure}{\arabic{figure}}
\setcounter{figure}{0}
\setcounter{table}{0}

\makeatletter
\renewcommand{\fnum@figure}{\textbf{Figure \thefigure}. }
\renewcommand{\fnum@table}{\textbf{Table \thetable. }}


\addcontentsline{toc}{section}{Abbreviation}
\section*{Abbreviation}
The following is the list of abbreviation utilized in the main paper.
\begin{enumerate}
    \item MOF: Metal-organic framework
    \item BBA: Building-block aware
    \item SE(3): Special Euclidean group in 3D space.
    \item SEM: Scanning electron microscopy.
    \item TGA: Thermogravimetric analysis.
    \item BET: Brunauer-Emmett-Teller.
\end{enumerate}

\clearpage
\begin{figure*}[t!]
    \includegraphics[width=0.4\textwidth]{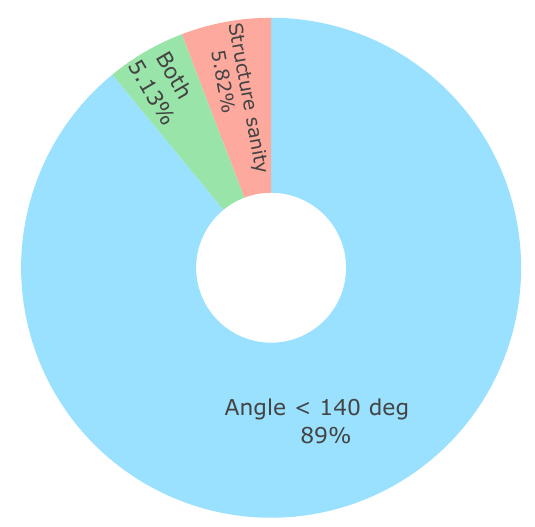}
    \caption{\textbf{Distribution for sources of invalidity in generated edges}. A structure is characterized as insane if it has nonphysical connectivity, such as floating atoms.}
    \label{Supp:edge_source_invalidty}
\end{figure*}

\begin{figure*}[t!]
    \includegraphics[width=0.4\textwidth]{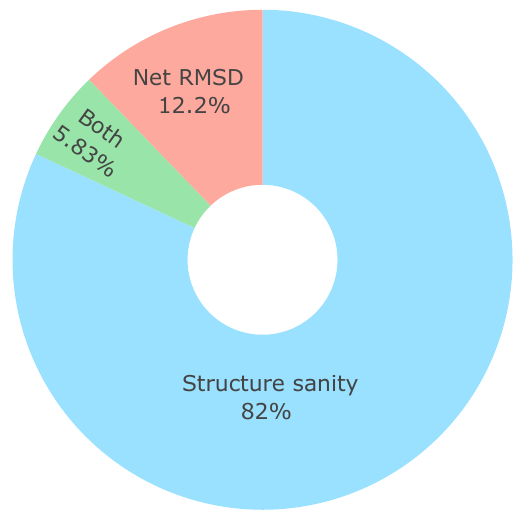}
    \caption{\textbf{Distribution for sources of invalidity in generated nodes}. A structure is characterized as insane if it has nonphysical connectivity, such as floating atoms.}
    \label{Supp:node_source_invalidty}
\end{figure*}

\begin{figure*}[t!]
    \includegraphics[width=0.8\textwidth]{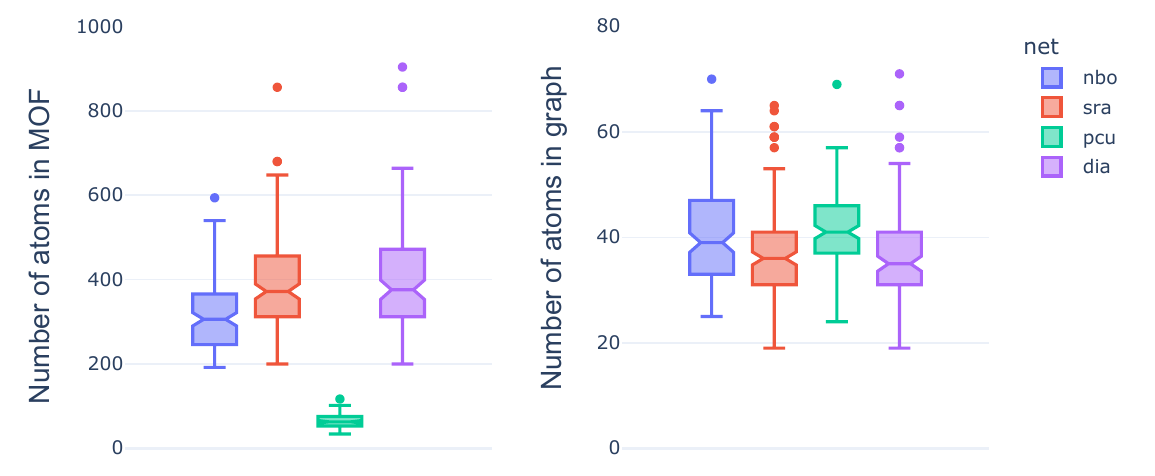}
    \caption{\textbf{Number of atoms for generated MOFs by size of MOFs (left) and size of graphs in scoring networks.}
    Nets are colored as the following: blue for \textit{nbo}, red for \textit{sra}, green for \textit{pcu}, and purple for \textit{dia}.
    }
    \label{Supp:natoms_mof_vs_graph}
\end{figure*}

\begin{figure}[htbp]
	\centering
	\begin{minipage}{0.49\linewidth}
		\centering
		\includegraphics[width=1\linewidth]{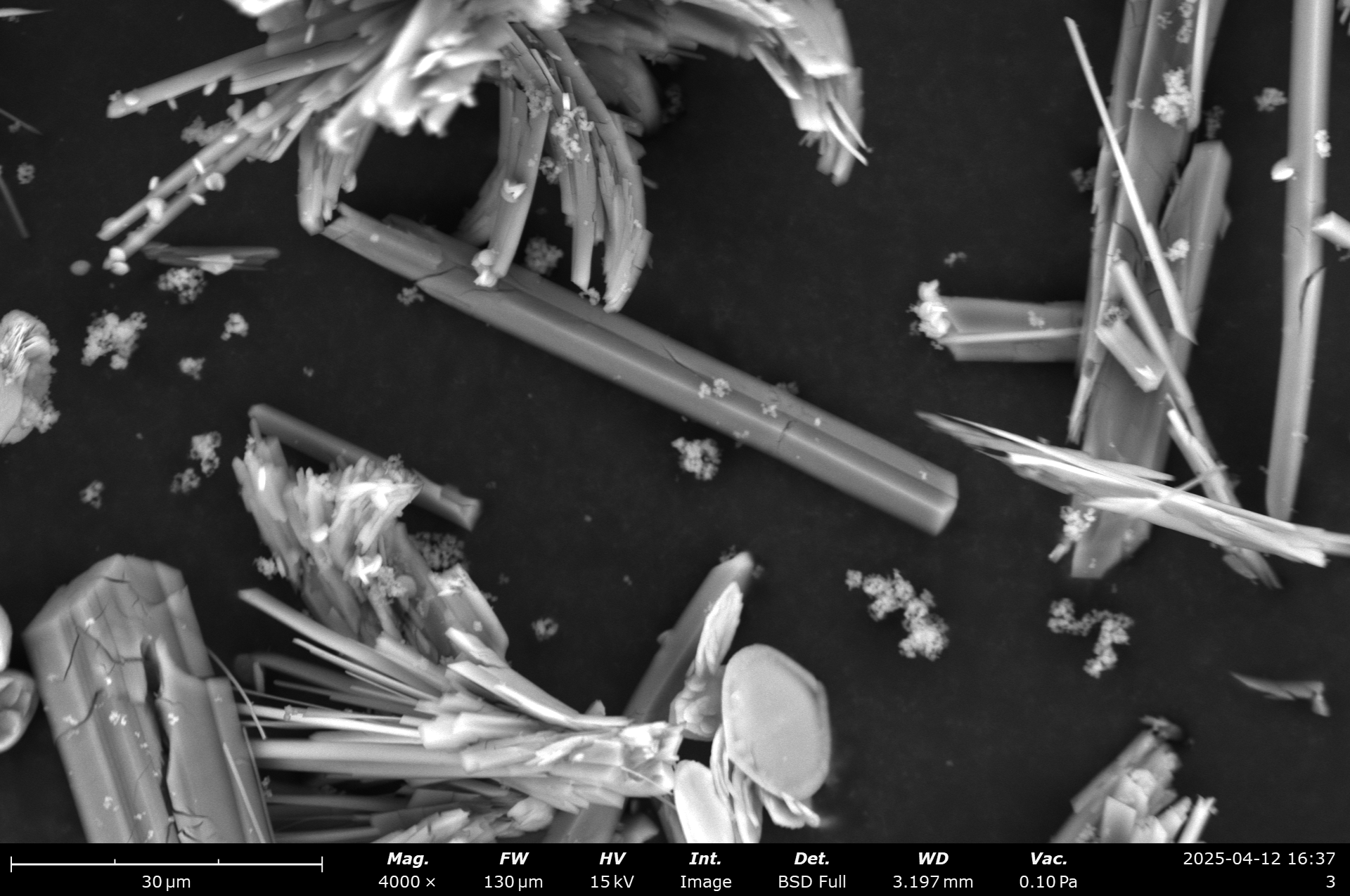}
	\end{minipage}
	\begin{minipage}{0.49\linewidth}
		\centering
		\includegraphics[width=1\linewidth]{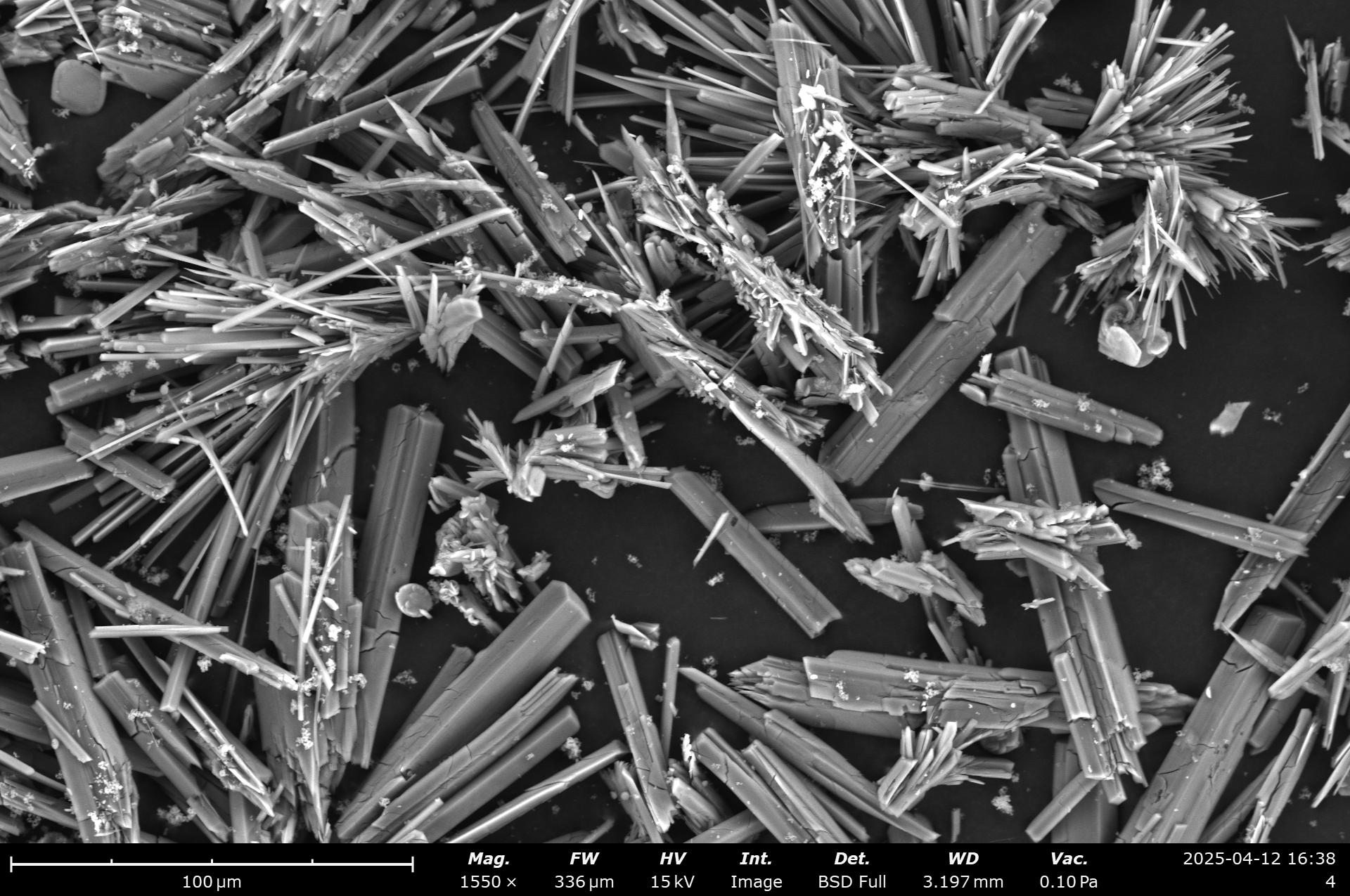}
	\end{minipage}
        \caption{
        SEM micrographs of $\mathrm{[Zn(1,4-TDC)(EtOH)_{2}]}$, which grow in the form of rods.
    }
     \label{Supp:sem}
\end{figure}

\begin{figure*}[t!]
    \includegraphics[width=0.7\textwidth]{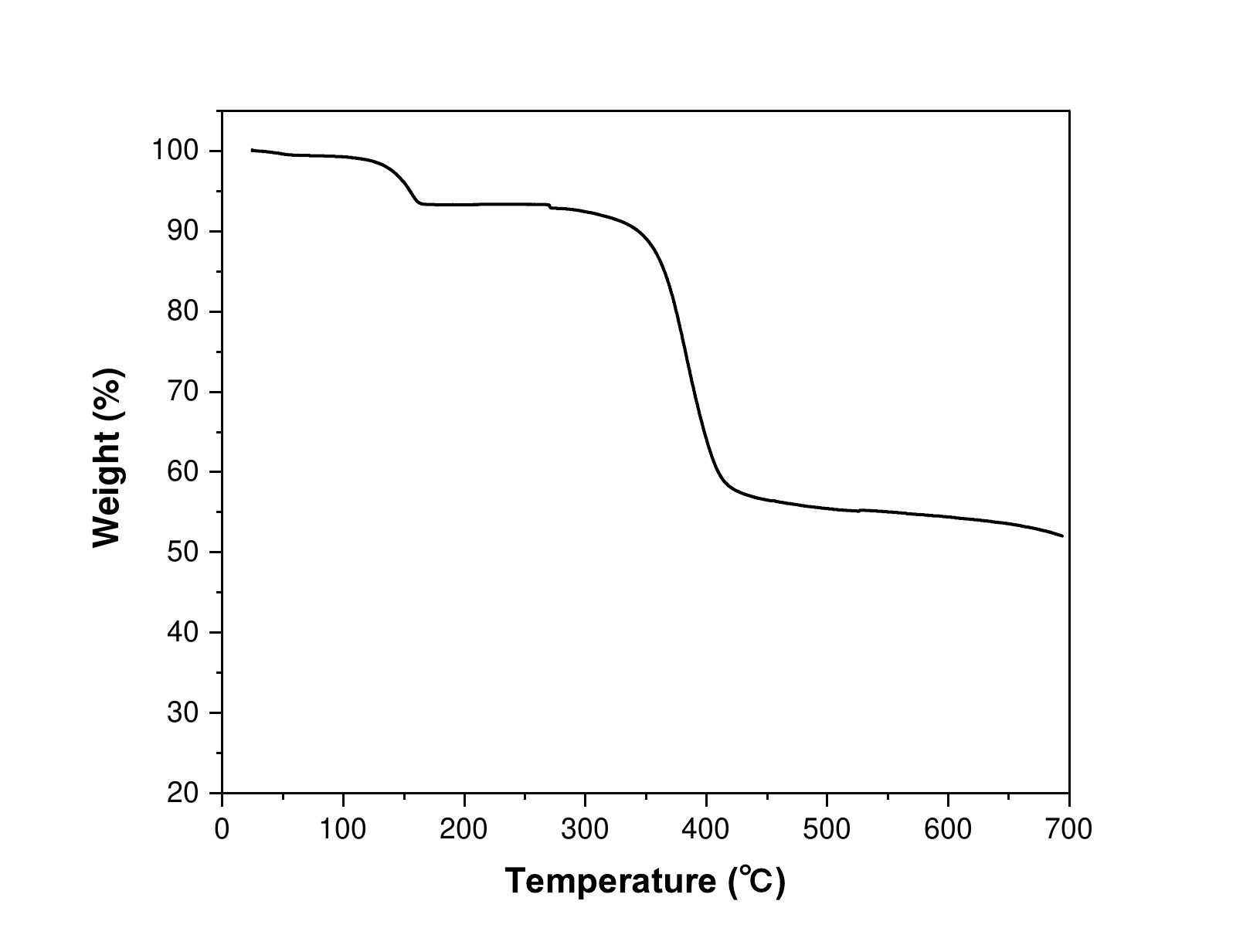}
    \caption{\textbf{TGA plot for $\mathrm{[Zn(1,4-TDC)(EtOH)_{2}]}$.}
TGA of the $\mathrm{[Zn(1,4-TDC)(EtOH)_{2}]}$ shows the quality changes of the two divisions. 
    }
    \label{Supp:tga}
\end{figure*}

\begin{figure*}[t!]
    \includegraphics[width=0.7\textwidth]{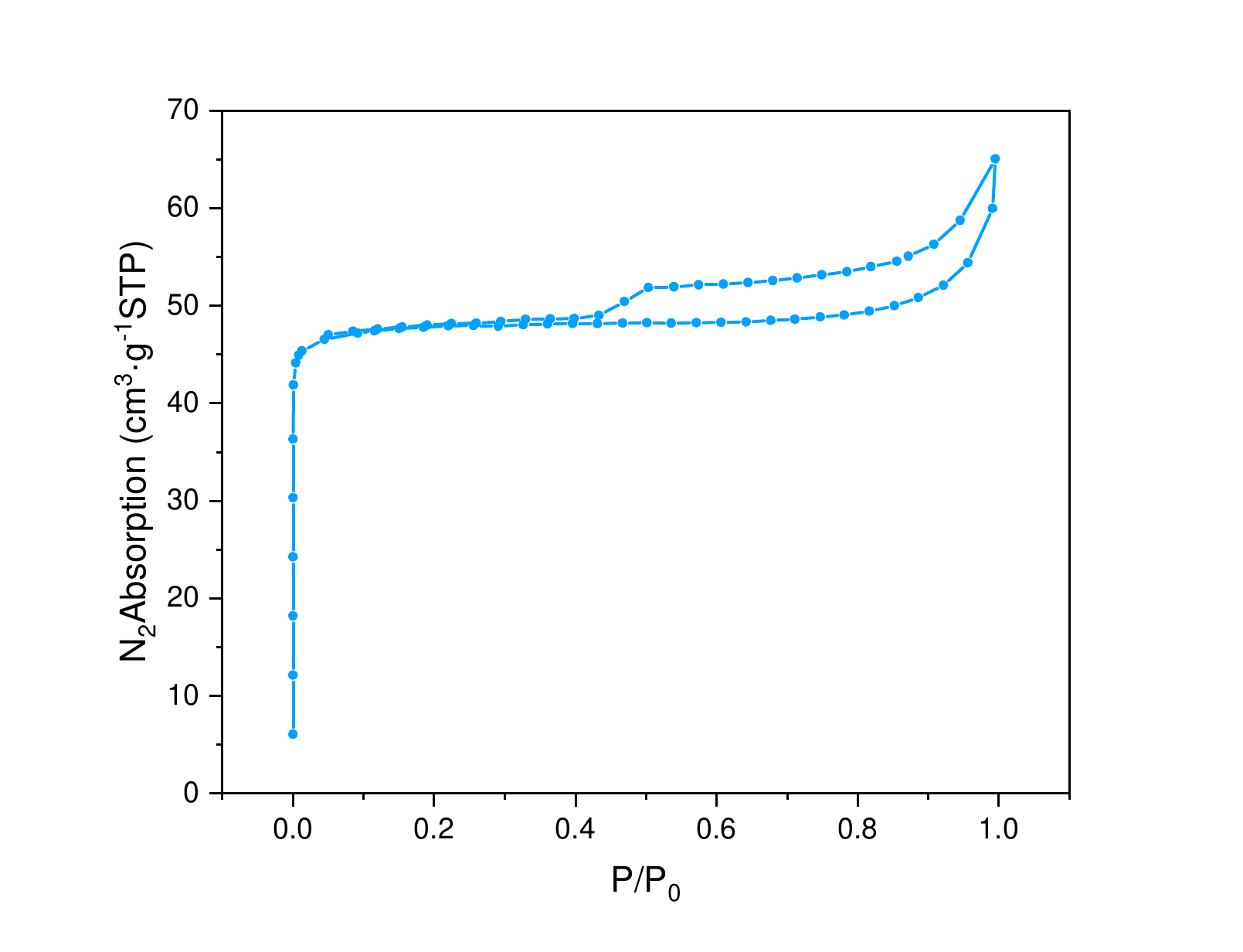}
    \caption{
Nitrogen adsorption isotherms for $\mathrm{[Zn(1,4-TDC)(EtOH)_{2}]}$ at 77K. The isotherm was fit to the BET equation  to give apparent BET surface areas of 140.37 ± 4.80 m²/g. 
    }
    \label{Supp:bet1}
\end{figure*}

\begin{figure*}[t!]
    \includegraphics[width=0.7\textwidth]{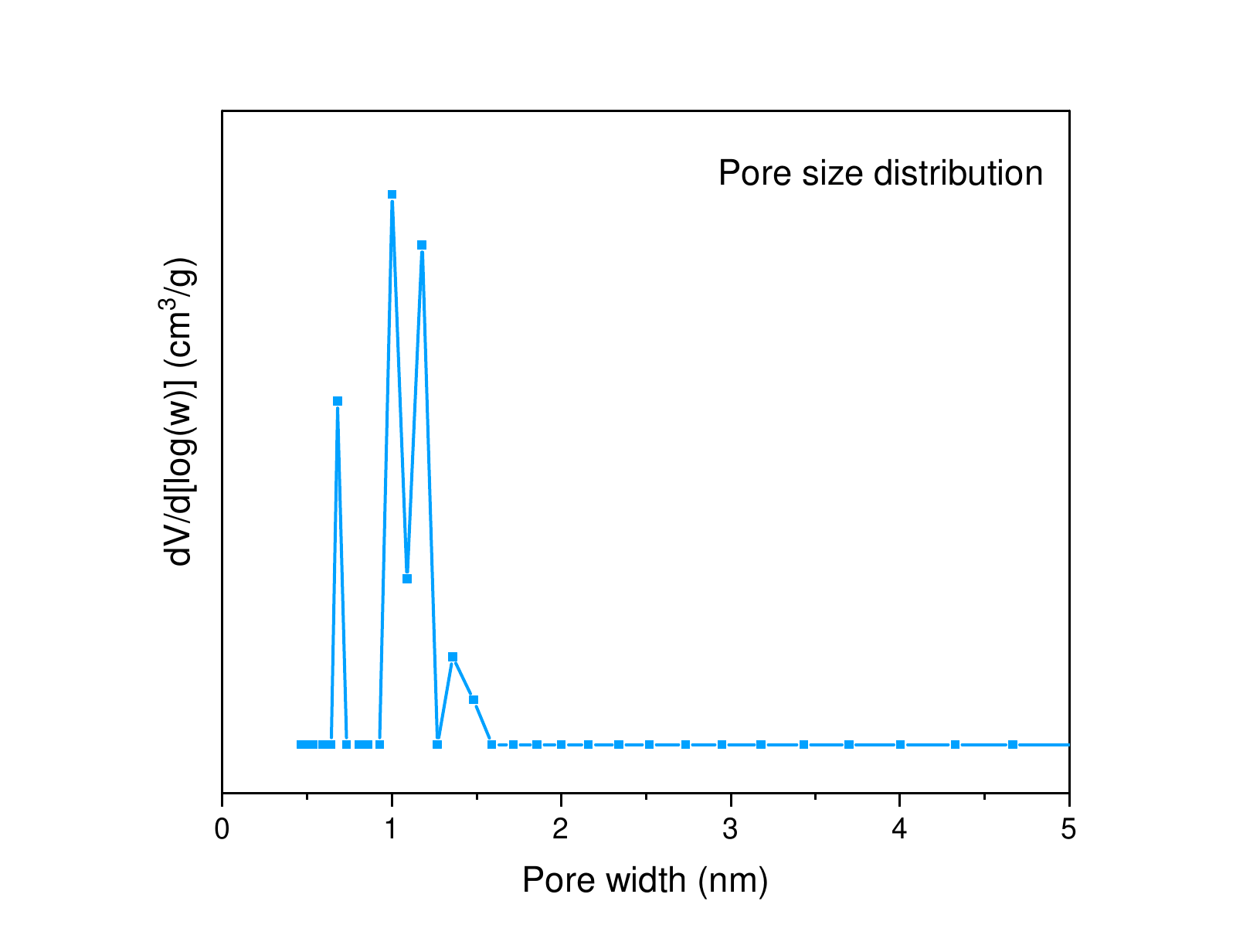}
    \caption{
    Barrett-Joyner-Halenda (BJH) pore size distribution curve using the Kruk-Jaroniec-Sayari  correction for hexagonal pores for $\mathrm{[Zn(1,4-TDC)(EtOH)_{2}]}$. Pore size: 1.4 nm.
    }
    \label{Supp:bet2}
\end{figure*}

%% file: main.bbl
\begin{thebibliography}{10}
\urlstyle{rm}
\expandafter\ifx\csname url\endcsname\relax
  \def\url#1{\texttt{#1}}\fi
\expandafter\ifx\csname urlprefix\endcsname\relax\def\urlprefix{URL }\fi
\expandafter\ifx\csname doiprefix\endcsname\relax\def\doiprefix{DOI: }\fi
\providecommand{\bibinfo}[2]{#2}
\providecommand{\eprint}[2][]{\url{#2}}

\bibitem{stein1993turning}
\bibinfo{author}{Stein, A.}, \bibinfo{author}{Keller, S.~W.} \& \bibinfo{author}{Mallouk, T.~E.}
\newblock \bibinfo{journal}{\bibinfo{title}{Turning down the heat: Design and mechanism in solid-state synthesis}}.
\newblock {\emph{\JournalTitle{Science}}} \textbf{\bibinfo{volume}{259}}, \bibinfo{pages}{1558--1564}, \doiprefix\url{10.1126/science.259.5101.1558} (\bibinfo{year}{1993}).

\bibitem{yaghi1998synthetic}
\bibinfo{author}{Yaghi, O.~M.}, \bibinfo{author}{Li, H.}, \bibinfo{author}{Davis, C.}, \bibinfo{author}{Richardson, D.} \& \bibinfo{author}{Groy, T.~L.}
\newblock \bibinfo{journal}{\bibinfo{title}{Synthetic strategies, structure patterns, and emerging properties in the chemistry of modular porous solids}}.
\newblock {\emph{\JournalTitle{Accounts of Chemical Research}}} \textbf{\bibinfo{volume}{31}}, \bibinfo{pages}{474--484} (\bibinfo{year}{1998}).

\bibitem{yaghi2000design}
\bibinfo{author}{Yaghi, O.}, \bibinfo{author}{O'Keeffe, M.} \& \bibinfo{author}{Kanatzidis, M.}
\newblock \bibinfo{journal}{\bibinfo{title}{Design of solids from molecular building blocks: golden opportunities for solid state chemistry}}.
\newblock {\emph{\JournalTitle{Journal of Solid State Chemistry}}} \textbf{\bibinfo{volume}{152}}, \bibinfo{pages}{1--2} (\bibinfo{year}{2000}).

\bibitem{yaghi2003reticular}
\bibinfo{author}{Yaghi, O.~M.}, \bibinfo{author}{O'Keeffe, M.}, \bibinfo{author}{Ockwig, N.~W.}, \bibinfo{author}{Chae, H.~K.}, \bibinfo{author}{Eddaoudi, M.} \& \bibinfo{author}{Kim, J.}
\newblock \bibinfo{journal}{\bibinfo{title}{Reticular synthesis and the design of new materials}}.
\newblock {\emph{\JournalTitle{Nature}}} \textbf{\bibinfo{volume}{423}}, \bibinfo{pages}{705--714} (\bibinfo{year}{2003}).

\bibitem{greed2025man}
\bibinfo{author}{Greed, S.} \& \bibinfo{author}{Yaghi, O.~M.}
\newblock \bibinfo{journal}{\bibinfo{title}{The man of mofs and more}}.
\newblock {\emph{\JournalTitle{Nature Reviews Chemistry}}} \bibinfo{pages}{1--3} (\bibinfo{year}{2025}).

\bibitem{dou2021atomically}
\bibinfo{author}{Dou, J.-H.}, \bibinfo{author}{Arguilla, M.~Q.}, \bibinfo{author}{Luo, Y.}, \bibinfo{author}{Li, J.}, \bibinfo{author}{Zhang, W.}, \bibinfo{author}{Sun, L.}, \bibinfo{author}{Mancuso, J.~L.}, \bibinfo{author}{Yang, L.}, \bibinfo{author}{Chen, T.}, \bibinfo{author}{Parent, L.~R.} \emph{et~al.}
\newblock \bibinfo{journal}{\bibinfo{title}{Atomically precise single-crystal structures of electrically conducting 2d metal--organic frameworks}}.
\newblock {\emph{\JournalTitle{Nature materials}}} \textbf{\bibinfo{volume}{20}}, \bibinfo{pages}{222--228}, \doiprefix\url{10.1038/s41563-020-00847-7} (\bibinfo{year}{2021}).

\bibitem{murray2009hydrogen}
\bibinfo{author}{Murray, L.~J.}, \bibinfo{author}{Dinc{\u{a}}, M.} \& \bibinfo{author}{Long, J.~R.}
\newblock \bibinfo{journal}{\bibinfo{title}{Hydrogen storage in metal--organic frameworks}}.
\newblock {\emph{\JournalTitle{Chemical Society Reviews}}} \textbf{\bibinfo{volume}{38}}, \bibinfo{pages}{1294--1314} (\bibinfo{year}{2009}).

\bibitem{Forse2020}
\bibinfo{author}{Forse, A.~C.}, \bibinfo{author}{Colwell, K.~A.}, \bibinfo{author}{Gonzalez, M.~I.}, \bibinfo{author}{Benders, S.}, \bibinfo{author}{Torres-Gavosto, R.~M.}, \bibinfo{author}{Blümich, B.}, \bibinfo{author}{Reimer, J.~A.} \& \bibinfo{author}{Long, J.~R.}
\newblock \bibinfo{journal}{\bibinfo{title}{Influence of pore size on carbon dioxide diffusion in two isoreticular metal–organic frameworks}}.
\newblock {\emph{\JournalTitle{Chemistry of Materials}}} \textbf{\bibinfo{volume}{32}}, \bibinfo{pages}{3570--3576}, \doiprefix\url{10.1021/acs.chemmater.0c00745} (\bibinfo{year}{2020}).

\bibitem{deng2010multiple}
\bibinfo{author}{Deng, H.}, \bibinfo{author}{Doonan, C.~J.}, \bibinfo{author}{Furukawa, H.}, \bibinfo{author}{Ferreira, R.~B.}, \bibinfo{author}{Towne, J.}, \bibinfo{author}{Knobler, C.~B.}, \bibinfo{author}{Wang, B.} \& \bibinfo{author}{Yaghi, O.~M.}
\newblock \bibinfo{journal}{\bibinfo{title}{Multiple functional groups of varying ratios in metal-organic frameworks}}.
\newblock {\emph{\JournalTitle{science}}} \textbf{\bibinfo{volume}{327}}, \bibinfo{pages}{846--850} (\bibinfo{year}{2010}).

\bibitem{eddaoudi2002systematic}
\bibinfo{author}{Eddaoudi, M.}, \bibinfo{author}{Kim, J.}, \bibinfo{author}{Rosi, N.}, \bibinfo{author}{Vodak, D.}, \bibinfo{author}{Wachter, J.}, \bibinfo{author}{O'Keeffe, M.} \& \bibinfo{author}{Yaghi, O.~M.}
\newblock \bibinfo{journal}{\bibinfo{title}{Systematic design of pore size and functionality in isoreticular mofs and their application in methane storage}}.
\newblock {\emph{\JournalTitle{Science}}} \textbf{\bibinfo{volume}{295}}, \bibinfo{pages}{469--472} (\bibinfo{year}{2002}).

\bibitem{lu2014tuning}
\bibinfo{author}{Lu, W.}, \bibinfo{author}{Wei, Z.}, \bibinfo{author}{Gu, Z.-Y.}, \bibinfo{author}{Liu, T.-F.}, \bibinfo{author}{Park, J.}, \bibinfo{author}{Park, J.}, \bibinfo{author}{Tian, J.}, \bibinfo{author}{Zhang, M.}, \bibinfo{author}{Zhang, Q.}, \bibinfo{author}{Gentle~III, T.} \emph{et~al.}
\newblock \bibinfo{journal}{\bibinfo{title}{Tuning the structure and function of metal--organic frameworks via linker design}}.
\newblock {\emph{\JournalTitle{Chemical Society Reviews}}} \textbf{\bibinfo{volume}{43}}, \bibinfo{pages}{5561--5593} (\bibinfo{year}{2014}).

\bibitem{furukawa2013chemistry}
\bibinfo{author}{Furukawa, H.}, \bibinfo{author}{Cordova, K.~E.}, \bibinfo{author}{O’Keeffe, M.} \& \bibinfo{author}{Yaghi, O.~M.}
\newblock \bibinfo{journal}{\bibinfo{title}{The chemistry and applications of metal-organic frameworks}}.
\newblock {\emph{\JournalTitle{Science}}} \textbf{\bibinfo{volume}{341}}, \bibinfo{pages}{1230444} (\bibinfo{year}{2013}).

\bibitem{chae2004route}
\bibinfo{author}{Chae, H.~K.}, \bibinfo{author}{Siberio-P{\'e}rez, D.~Y.}, \bibinfo{author}{Kim, J.}, \bibinfo{author}{Go, Y.}, \bibinfo{author}{Eddaoudi, M.}, \bibinfo{author}{Matzger, A.~J.}, \bibinfo{author}{O'keeffe, M.}, \bibinfo{author}{Yaghi, O.~M.}, \bibinfo{author}{Design, M.} \& \bibinfo{author}{Group, D.}
\newblock \bibinfo{journal}{\bibinfo{title}{A route to high surface area, porosity and inclusion of large molecules in crystals}}.
\newblock {\emph{\JournalTitle{Nature}}} \textbf{\bibinfo{volume}{427}}, \bibinfo{pages}{523--527} (\bibinfo{year}{2004}).

\bibitem{baumann2019metal}
\bibinfo{author}{Baumann, A.~E.}, \bibinfo{author}{Burns, D.~A.}, \bibinfo{author}{Liu, B.} \& \bibinfo{author}{Thoi, V.~S.}
\newblock \bibinfo{journal}{\bibinfo{title}{Metal-organic framework functionalization and design strategies for advanced electrochemical energy storage devices}}.
\newblock {\emph{\JournalTitle{Communications Chemistry}}} \textbf{\bibinfo{volume}{2}}, \bibinfo{pages}{86} (\bibinfo{year}{2019}).

\bibitem{yuan2018stable}
\bibinfo{author}{Yuan, S.}, \bibinfo{author}{Feng, L.}, \bibinfo{author}{Wang, K.}, \bibinfo{author}{Pang, J.}, \bibinfo{author}{Bosch, M.}, \bibinfo{author}{Lollar, C.}, \bibinfo{author}{Sun, Y.}, \bibinfo{author}{Qin, J.}, \bibinfo{author}{Yang, X.}, \bibinfo{author}{Zhang, P.} \emph{et~al.}
\newblock \bibinfo{journal}{\bibinfo{title}{Stable metal--organic frameworks: design, synthesis, and applications}}.
\newblock {\emph{\JournalTitle{Advanced Materials}}} \textbf{\bibinfo{volume}{30}}, \bibinfo{pages}{1704303} (\bibinfo{year}{2018}).

\bibitem{sheberla2017conductive}
\bibinfo{author}{Sheberla, D.}, \bibinfo{author}{Bachman, J.~C.}, \bibinfo{author}{Elias, J.~S.}, \bibinfo{author}{Sun, C.-J.}, \bibinfo{author}{Shao-Horn, Y.} \& \bibinfo{author}{Dinc{\u{a}}, M.}
\newblock \bibinfo{journal}{\bibinfo{title}{Conductive mof electrodes for stable supercapacitors with high areal capacitance}}.
\newblock {\emph{\JournalTitle{Nature materials}}} \textbf{\bibinfo{volume}{16}}, \bibinfo{pages}{220--224} (\bibinfo{year}{2017}).

\bibitem{niu2022conductive}
\bibinfo{author}{Niu, L.}, \bibinfo{author}{Wu, T.}, \bibinfo{author}{Chen, M.}, \bibinfo{author}{Yang, L.}, \bibinfo{author}{Yang, J.}, \bibinfo{author}{Wang, Z.}, \bibinfo{author}{Kornyshev, A.~A.}, \bibinfo{author}{Jiang, H.}, \bibinfo{author}{Bi, S.} \& \bibinfo{author}{Feng, G.}
\newblock \bibinfo{journal}{\bibinfo{title}{Conductive metal--organic frameworks for supercapacitors}}.
\newblock {\emph{\JournalTitle{Advanced Materials}}} \textbf{\bibinfo{volume}{34}}, \bibinfo{pages}{2200999} (\bibinfo{year}{2022}).

\bibitem{yang2024oxygen}
\bibinfo{author}{Yang, J.}, \bibinfo{author}{Xu, H.}, \bibinfo{author}{Li, J.}, \bibinfo{author}{Gong, K.}, \bibinfo{author}{Yue, F.}, \bibinfo{author}{Han, X.}, \bibinfo{author}{Wu, K.}, \bibinfo{author}{Shao, P.}, \bibinfo{author}{Fu, Q.}, \bibinfo{author}{Zhu, Y.} \emph{et~al.}
\newblock \bibinfo{journal}{\bibinfo{title}{Oxygen-and proton-transporting open framework ionomer for medium-temperature fuel cells}}.
\newblock {\emph{\JournalTitle{Science}}} \textbf{\bibinfo{volume}{385}}, \bibinfo{pages}{1115--1120} (\bibinfo{year}{2024}).

\bibitem{stanley2024analysis}
\bibinfo{author}{Stanley, P.}, \bibinfo{author}{Ramm, V.}, \bibinfo{author}{Fischer, R.} \& \bibinfo{author}{Warnan, J.}
\newblock \bibinfo{journal}{\bibinfo{title}{Analysis of metal--organic framework-based photosynthetic co2 reduction}}.
\newblock {\emph{\JournalTitle{Nature Synthesis}}} \textbf{\bibinfo{volume}{3}}, \bibinfo{pages}{307--318} (\bibinfo{year}{2024}).

\bibitem{fu2012amine}
\bibinfo{author}{Fu, Y.}, \bibinfo{author}{Sun, D.}, \bibinfo{author}{Chen, Y.}, \bibinfo{author}{Huang, R.}, \bibinfo{author}{Ding, Z.}, \bibinfo{author}{Fu, X.} \& \bibinfo{author}{Li, Z.}
\newblock \bibinfo{journal}{\bibinfo{title}{An amine-functionalized titanium metal-organic framework photocatalyst with visible-light-induced activity for co2 reduction.}}
\newblock {\emph{\JournalTitle{Angewandte Chemie (International ed. in English)}}} \textbf{\bibinfo{volume}{51}}, \bibinfo{pages}{3364--3367} (\bibinfo{year}{2012}).

\bibitem{niu2019promoting}
\bibinfo{author}{Niu, Z.}, \bibinfo{author}{Zhang, W.}, \bibinfo{author}{Lan, P.~C.}, \bibinfo{author}{Aguila, B.} \& \bibinfo{author}{Ma, S.}
\newblock \bibinfo{journal}{\bibinfo{title}{Promoting frustrated lewis pairs for heterogeneous chemoselective hydrogenation via the tailored pore environment within metal--organic frameworks}}.
\newblock {\emph{\JournalTitle{Angewandte Chemie International Edition}}} \textbf{\bibinfo{volume}{58}}, \bibinfo{pages}{7420--7424} (\bibinfo{year}{2019}).

\bibitem{niu2018metal}
\bibinfo{author}{Niu, Z.}, \bibinfo{author}{Gunatilleke, W. D.~B.}, \bibinfo{author}{Sun, Q.}, \bibinfo{author}{Lan, P.~C.}, \bibinfo{author}{Perman, J.}, \bibinfo{author}{Ma, J.-G.}, \bibinfo{author}{Cheng, Y.}, \bibinfo{author}{Aguila, B.} \& \bibinfo{author}{Ma, S.}
\newblock \bibinfo{journal}{\bibinfo{title}{Metal-organic framework anchored with a lewis pair as a new paradigm for catalysis}}.
\newblock {\emph{\JournalTitle{Chem}}} \textbf{\bibinfo{volume}{4}}, \bibinfo{pages}{2587--2599} (\bibinfo{year}{2018}).

\bibitem{kreno2012metal}
\bibinfo{author}{Kreno, L.~E.}, \bibinfo{author}{Leong, K.}, \bibinfo{author}{Farha, O.~K.}, \bibinfo{author}{Allendorf, M.}, \bibinfo{author}{Van~Duyne, R.~P.} \& \bibinfo{author}{Hupp, J.~T.}
\newblock \bibinfo{journal}{\bibinfo{title}{Metal--organic framework materials as chemical sensors}}.
\newblock {\emph{\JournalTitle{Chemical reviews}}} \textbf{\bibinfo{volume}{112}}, \bibinfo{pages}{1105--1125} (\bibinfo{year}{2012}).

\bibitem{allendorf2009luminescent}
\bibinfo{author}{Allendorf, M.~D.}, \bibinfo{author}{Bauer, C.~A.}, \bibinfo{author}{Bhakta, R.} \& \bibinfo{author}{Houk, R.}
\newblock \bibinfo{journal}{\bibinfo{title}{Luminescent metal--organic frameworks}}.
\newblock {\emph{\JournalTitle{Chemical Society Reviews}}} \textbf{\bibinfo{volume}{38}}, \bibinfo{pages}{1330--1352} (\bibinfo{year}{2009}).

\bibitem{jin2024metal}
\bibinfo{author}{Jin, Y.}, \bibinfo{author}{Liu, H.}, \bibinfo{author}{Feng, M.}, \bibinfo{author}{Ma, Q.} \& \bibinfo{author}{Wang, B.}
\newblock \bibinfo{journal}{\bibinfo{title}{Metal-organic frameworks for air pollution purification and detection}}.
\newblock {\emph{\JournalTitle{Advanced Functional Materials}}} \textbf{\bibinfo{volume}{34}}, \bibinfo{pages}{2304773} (\bibinfo{year}{2024}).

\bibitem{li2011carbon}
\bibinfo{author}{Li, J.-R.}, \bibinfo{author}{Ma, Y.}, \bibinfo{author}{McCarthy, M.~C.}, \bibinfo{author}{Sculley, J.}, \bibinfo{author}{Yu, J.}, \bibinfo{author}{Jeong, H.-K.}, \bibinfo{author}{Balbuena, P.~B.} \& \bibinfo{author}{Zhou, H.-C.}
\newblock \bibinfo{journal}{\bibinfo{title}{Carbon dioxide capture-related gas adsorption and separation in metal-organic frameworks}}.
\newblock {\emph{\JournalTitle{Coordination Chemistry Reviews}}} \textbf{\bibinfo{volume}{255}}, \bibinfo{pages}{1791--1823} (\bibinfo{year}{2011}).

\bibitem{CSDGroom}
\bibinfo{author}{Groom, C.~R.}, \bibinfo{author}{Bruno, I.~J.}, \bibinfo{author}{Lightfoot, M.~P.} \& \bibinfo{author}{Ward, S.~C.}
\newblock \bibinfo{journal}{\bibinfo{title}{The cambridge structural database}}.
\newblock {\emph{\JournalTitle{Acta Crystallographica Section B Structural Science, Crystal Engineering and Materials}}} \textbf{\bibinfo{volume}{72}}, \bibinfo{pages}{171–179}, \doiprefix\url{10.1107/s2052520616003954} (\bibinfo{year}{2016}).

\bibitem{Chung2019}
\bibinfo{author}{Chung, Y.~G.}, \bibinfo{author}{Haldoupis, E.}, \bibinfo{author}{Bucior, B.~J.}, \bibinfo{author}{Haranczyk, M.}, \bibinfo{author}{Lee, S.}, \bibinfo{author}{Zhang, H.}, \bibinfo{author}{Vogiatzis, K.~D.}, \bibinfo{author}{Milisavljevic, M.}, \bibinfo{author}{Ling, S.}, \bibinfo{author}{Camp, J.~S.}, \bibinfo{author}{Slater, B.}, \bibinfo{author}{Siepmann, J.~I.}, \bibinfo{author}{Sholl, D.~S.} \& \bibinfo{author}{Snurr, R.~Q.}
\newblock \bibinfo{journal}{\bibinfo{title}{Advances, updates, and analytics for the computation-ready, experimental metal–organic framework database: Core mof 2019}}.
\newblock {\emph{\JournalTitle{Journal of Chemical \& Engineering Data}}} \textbf{\bibinfo{volume}{64}}, \bibinfo{pages}{5985--5998}, \doiprefix\url{10.1021/acs.jced.9b00835} (\bibinfo{year}{2019}).

\bibitem{Zhao2024_CoRE_MOF_DB}
\bibinfo{author}{Zhao, G.}, \bibinfo{author}{Brabson, L.}, \bibinfo{author}{Chheda, S.}, \bibinfo{author}{Huang, J.}, \bibinfo{author}{Kim, H.}, \bibinfo{author}{Liu, K.} \emph{et~al.}
\newblock \bibinfo{title}{{CoRE MOF DB}: a curated experimental metal-organic framework database with machine-learned properties for integrated material–process screening}.
\newblock \bibinfo{howpublished}{ChemRxiv preprint}, \doiprefix\url{10.26434/chemrxiv-2024-nvmnr} (\bibinfo{year}{2024}).
\newblock \bibinfo{note}{Preprint; not peer-reviewed}.

\bibitem{CSDMOFFairen2017}
\bibinfo{author}{Moghadam, P.~Z.}, \bibinfo{author}{Li, A.}, \bibinfo{author}{Wiggin, S.~B.}, \bibinfo{author}{Tao, A.}, \bibinfo{author}{Maloney, A. G.~P.}, \bibinfo{author}{Wood, P.~A.}, \bibinfo{author}{Ward, S.~C.} \& \bibinfo{author}{Fairen-Jimenez, D.}
\newblock \bibinfo{journal}{\bibinfo{title}{Development of a cambridge structural database subset: A collection of metal–organic frameworks for past, present, and future}}.
\newblock {\emph{\JournalTitle{Chemistry of Materials}}} \textbf{\bibinfo{volume}{29}}, \bibinfo{pages}{2618--2625}, \doiprefix\url{10.1021/acs.chemmater.7b00441} (\bibinfo{year}{2017}).

\bibitem{Wilmer2012}
\bibinfo{author}{Wilmer, C.~E.}, \bibinfo{author}{Leaf, M.}, \bibinfo{author}{Lee, C.~Y.}, \bibinfo{author}{Farha, O.~K.}, \bibinfo{author}{Hauser, B.~G.}, \bibinfo{author}{Hupp, J.~T.} \& \bibinfo{author}{Snurr, R.~Q.}
\newblock \bibinfo{journal}{\bibinfo{title}{Large-scale screening of hypothetical metal–organic frameworks}}.
\newblock {\emph{\JournalTitle{Nature Chemistry}}} \textbf{\bibinfo{volume}{4}}, \bibinfo{pages}{83--89}, \doiprefix\url{10.1038/nchem.1192} (\bibinfo{year}{2012}).

\bibitem{BoydWooNature}
\bibinfo{author}{Boyd, P.~G.}, \bibinfo{author}{Chidambaram, A.}, \bibinfo{author}{García-Díez, E.}, \bibinfo{author}{Ireland, C.~P.}, \bibinfo{author}{Daff, T.~D.}, \bibinfo{author}{Bounds, R.}, \bibinfo{author}{Gładysiak, A.}, \bibinfo{author}{Schouwink, P.}, \bibinfo{author}{Moosavi, S.~M.}, \bibinfo{author}{Maroto-Valer, M.~M.}, \bibinfo{author}{Reimer, J.~A.}, \bibinfo{author}{Navarro, J. A.~R.}, \bibinfo{author}{Woo, T.~K.}, \bibinfo{author}{Garcia, S.}, \bibinfo{author}{Stylianou, K.~C.} \& \bibinfo{author}{Smit, B.}
\newblock \bibinfo{journal}{\bibinfo{title}{Data-driven design of metal–organic frameworks for wet flue gas co2 capture}}.
\newblock {\emph{\JournalTitle{Nature}}} \textbf{\bibinfo{volume}{576}}, \bibinfo{pages}{253--256}, \doiprefix\url{10.1038/s41586-019-1798-7} (\bibinfo{year}{2019}).

\bibitem{ColonTobacco2017}
\bibinfo{author}{Colón, Y.~J.}, \bibinfo{author}{Gómez-Gualdrón, D.~A.} \& \bibinfo{author}{Snurr, R.~Q.}
\newblock \bibinfo{journal}{\bibinfo{title}{Topologically guided, automated construction of metal–organic frameworks and their evaluation for energy-related applications}}.
\newblock {\emph{\JournalTitle{Crystal Growth \& Design}}} \textbf{\bibinfo{volume}{17}}, \bibinfo{pages}{5801–5810}, \doiprefix\url{10.1021/acs.cgd.7b00848} (\bibinfo{year}{2017}).

\bibitem{Nandy2023Matter}
\bibinfo{author}{Nandy, A.}, \bibinfo{author}{Yue, S.}, \bibinfo{author}{Oh, C.}, \bibinfo{author}{Duan, C.}, \bibinfo{author}{Terrones, G.~G.}, \bibinfo{author}{Chung, Y.~G.} \& \bibinfo{author}{Kulik, H.~J.}
\newblock \bibinfo{journal}{\bibinfo{title}{A database of ultrastable mofs reassembled from stable fragments with machine learning models}}.
\newblock {\emph{\JournalTitle{Matter}}} \textbf{\bibinfo{volume}{6}}, \bibinfo{pages}{1585--1603}, \doiprefix\url{10.1016/j.matt.2023.03.009} (\bibinfo{year}{2023}).

\bibitem{ARCMOF}
\bibinfo{author}{Burner, J.}, \bibinfo{author}{Luo, J.}, \bibinfo{author}{White, A.}, \bibinfo{author}{Mirmiran, A.}, \bibinfo{author}{Kwon, O.}, \bibinfo{author}{Boyd, P.~G.}, \bibinfo{author}{Maley, S.}, \bibinfo{author}{Gibaldi, M.}, \bibinfo{author}{Simrod, S.}, \bibinfo{author}{Ogden, V.} \& \bibinfo{author}{Woo, T.~K.}
\newblock \bibinfo{journal}{\bibinfo{title}{Arc–mof: A diverse database of metal-organic frameworks with dft-derived partial atomic charges and descriptors for machine learning}}.
\newblock {\emph{\JournalTitle{Chemistry of Materials}}} \textbf{\bibinfo{volume}{35}}, \bibinfo{pages}{900--916}, \doiprefix\url{10.1021/acs.chemmater.2c02485} (\bibinfo{year}{2023}).

\bibitem{MajumdarDiversifyingACSAMI}
\bibinfo{author}{Majumdar, S.}, \bibinfo{author}{Moosavi, S.~M.}, \bibinfo{author}{Jablonka, K.~M.}, \bibinfo{author}{Ongari, D.} \& \bibinfo{author}{Smit, B.}
\newblock \bibinfo{journal}{\bibinfo{title}{Diversifying databases of metal organic frameworks for high-throughput computational screening.}}
\newblock {\emph{\JournalTitle{ACS Applied Materials \& Interfaces}}} \textbf{\bibinfo{volume}{3}}, \bibinfo{pages}{61004–61014}, \doiprefix\url{10.1021/acsami.1c16220} (\bibinfo{year}{2021}).

\bibitem{MoosaviDiversity2020}
\bibinfo{author}{Moosavi, S.~M.}, \bibinfo{author}{Nandy, A.}, \bibinfo{author}{Jablonka, K.~M.}, \bibinfo{author}{Ongari, D.}, \bibinfo{author}{Janet, J.~P.}, \bibinfo{author}{Boyd, P.~G.}, \bibinfo{author}{Lee, Y.}, \bibinfo{author}{Smit, B.} \& \bibinfo{author}{Kulik, H.~J.}
\newblock \bibinfo{journal}{\bibinfo{title}{Understanding the diversity of the metal-organic framework ecosystem.}}
\newblock {\emph{\JournalTitle{Nature Communications}}} \bibinfo{pages}{4068}, \doiprefix\url{10.1038/s41467-020-17755-8} (\bibinfo{year}{2020}).

\bibitem{Fu2023}
\bibinfo{author}{Fu, Y.}, \bibinfo{author}{Yao, Y.}, \bibinfo{author}{Forse, A.~C.}, \bibinfo{author}{Li, J.}, \bibinfo{author}{Mochizuki, K.}, \bibinfo{author}{Long, J.~R.}, \bibinfo{author}{Reimer, J.~A.}, \bibinfo{author}{De~Pa{\"e}pe, G.} \& \bibinfo{author}{Kong, X.}
\newblock \bibinfo{journal}{\bibinfo{title}{Solvent-derived defects suppress adsorption in mof-74}}.
\newblock {\emph{\JournalTitle{Nature Communications}}} \textbf{\bibinfo{volume}{14}}, \bibinfo{pages}{2386}, \doiprefix\url{10.1038/s41467-023-38155-8} (\bibinfo{year}{2023}).

\bibitem{Lee2015}
\bibinfo{author}{Lee, K.}, \bibinfo{author}{Howe, J.~D.}, \bibinfo{author}{Lin, L.-C.}, \bibinfo{author}{Smit, B.} \& \bibinfo{author}{Neaton, J.~B.}
\newblock \bibinfo{journal}{\bibinfo{title}{Small-molecule adsorption in open-site metal--organic frameworks: A systematic density functional theory study for rational design}}.
\newblock {\emph{\JournalTitle{Chemistry of Materials}}} \textbf{\bibinfo{volume}{27}}, \bibinfo{pages}{668--678}, \doiprefix\url{10.1021/cm502760q} (\bibinfo{year}{2015}).

\bibitem{Nazarian2017}
\bibinfo{author}{Nazarian, D.}, \bibinfo{author}{Camp, J.~S.}, \bibinfo{author}{Chung, Y.~G.}, \bibinfo{author}{Snurr, R.~Q.} \& \bibinfo{author}{Sholl, D.~S.}
\newblock \bibinfo{journal}{\bibinfo{title}{Large-scale refinement of metal-organic framework structures using density functional theory}}.
\newblock {\emph{\JournalTitle{Chemistry of Materials}}} \textbf{\bibinfo{volume}{29}}, \bibinfo{pages}{2521--2528}, \doiprefix\url{10.1021/acs.chemmater.6b04226} (\bibinfo{year}{2017}).

\bibitem{shengchao2023}
\bibinfo{author}{Liu, S.}, \bibinfo{author}{Du, w.}, \bibinfo{author}{Li, Y.}, \bibinfo{author}{Li, Z.}, \bibinfo{author}{Zheng, Z.}, \bibinfo{author}{Duan, C.}, \bibinfo{author}{Ma, Z.-M.}, \bibinfo{author}{Yaghi, O.}, \bibinfo{author}{Anandkumar, A.}, \bibinfo{author}{Borgs, C.}, \bibinfo{author}{Chayes, J.}, \bibinfo{author}{Guo, H.} \& \bibinfo{author}{Tang, J.}
\newblock \bibinfo{title}{Symmetry-informed geometric representation for molecules, proteins, and crystalline materials}.
\newblock In \bibinfo{editor}{Oh, A.}, \bibinfo{editor}{Naumann, T.}, \bibinfo{editor}{Globerson, A.}, \bibinfo{editor}{Saenko, K.}, \bibinfo{editor}{Hardt, M.} \& \bibinfo{editor}{Levine, S.} (eds.) \emph{\bibinfo{booktitle}{Advances in Neural Information Processing Systems}}, vol.~\bibinfo{volume}{36}, \bibinfo{pages}{66084--66101} (\bibinfo{publisher}{Curran Associates, Inc.}, \bibinfo{year}{2023}).

\bibitem{Zhao2024PACMAN}
\bibinfo{author}{Zhao, G.} \& \bibinfo{author}{Chung, Y.~G.}
\newblock \bibinfo{journal}{\bibinfo{title}{Pacman: A robust partial atomic charge predicter for nanoporous materials based on crystal graph convolution networks}}.
\newblock {\emph{\JournalTitle{Journal of Chemical Theory and Computation}}} \textbf{\bibinfo{volume}{20}}, \bibinfo{pages}{5368--5380}, \doiprefix\url{10.1021/acs.jctc.4c00434} (\bibinfo{year}{2024}).

\bibitem{Altintas2021}
\bibinfo{author}{Altintas, C.}, \bibinfo{author}{Altundal, O.~F.}, \bibinfo{author}{Keskin, S.} \& \bibinfo{author}{Yildirim, R.}
\newblock \bibinfo{journal}{\bibinfo{title}{Machine learning meets with metal organic frameworks for gas storage and separation}}.
\newblock {\emph{\JournalTitle{Journal of Chemical Information and Modeling}}} \textbf{\bibinfo{volume}{61}}, \bibinfo{pages}{2131--2146}, \doiprefix\url{10.1021/acs.jcim.1c00191} (\bibinfo{year}{2021}).

\bibitem{Rosen2022}
\bibinfo{author}{Rosen, A.~S.}, \bibinfo{author}{Fung, V.}, \bibinfo{author}{Huck, P.}, \bibinfo{author}{O'Donnell, C.~T.}, \bibinfo{author}{Horton, M.~K.}, \bibinfo{author}{Truhlar, D.~G.}, \bibinfo{author}{Persson, K.~A.}, \bibinfo{author}{Notestein, J.~M.} \& \bibinfo{author}{Snurr, R.~Q.}
\newblock \bibinfo{journal}{\bibinfo{title}{High-throughput predictions of metal--organic framework electronic properties: theoretical challenges, graph neural networks, and data exploration}}.
\newblock {\emph{\JournalTitle{npj Computational Materials}}} \textbf{\bibinfo{volume}{8}}, \bibinfo{pages}{112}, \doiprefix\url{10.1038/s41524-022-00796-6} (\bibinfo{year}{2022}).

\bibitem{Dou2024}
\bibinfo{author}{Lin, J.}, \bibinfo{author}{Zhang, H.}, \bibinfo{author}{Asadi, M.}, \bibinfo{author}{Zhao, K.}, \bibinfo{author}{Yang, L.}, \bibinfo{author}{Fan, Y.}, \bibinfo{author}{Zhu, J.}, \bibinfo{author}{Liu, Q.}, \bibinfo{author}{Sun, L.}, \bibinfo{author}{Xie, W.~J.}, \bibinfo{author}{Duan, C.}, \bibinfo{author}{Mo, F.} \& \bibinfo{author}{Dou, J.-H.}
\newblock \bibinfo{journal}{\bibinfo{title}{Machine learning-driven discovery and structure--activity relationship analysis of conductive metal--organic frameworks}}.
\newblock {\emph{\JournalTitle{Chemistry of Materials}}} \textbf{\bibinfo{volume}{36}}, \bibinfo{pages}{5436--5445}, \doiprefix\url{10.1021/acs.chemmater.4c00229} (\bibinfo{year}{2024}).

\bibitem{Moghadam2019}
\bibinfo{author}{Moghadam, P.~Z.}, \bibinfo{author}{Rogge, S.~M.}, \bibinfo{author}{Li, A.}, \bibinfo{author}{Chow, C.-M.}, \bibinfo{author}{Wieme, J.}, \bibinfo{author}{Moharrami, N.}, \bibinfo{author}{Aragones-Anglada, M.}, \bibinfo{author}{Conduit, G.}, \bibinfo{author}{Gomez-Gualdron, D.~A.}, \bibinfo{author}{Van~Speybroeck, V.} \& \bibinfo{author}{Fairen-Jimenez, D.}
\newblock \bibinfo{journal}{\bibinfo{title}{Structure-mechanical stability relations of metal-organic frameworks via machine learning}}.
\newblock {\emph{\JournalTitle{Matter}}} \textbf{\bibinfo{volume}{1}}, \bibinfo{pages}{219--234}, \doiprefix\url{10.1016/j.matt.2019.03.002} (\bibinfo{year}{2019}).

\bibitem{NandyJACS2021}
\bibinfo{author}{Nandy, A.}, \bibinfo{author}{Terrones, G.}, \bibinfo{author}{Arunachalam, N.}, \bibinfo{author}{Duan, C.}, \bibinfo{author}{Kastner, D.~W.} \& \bibinfo{author}{Kulik, H.~J.}
\newblock \bibinfo{journal}{\bibinfo{title}{Mofsimplify, machine learning models with extracted stability data of three thousand metal–organic frameworks}}.
\newblock {\emph{\JournalTitle{Scientific Data}}} \textbf{\bibinfo{volume}{9}}, \bibinfo{pages}{74}, \doiprefix\url{10.1038/s41597-022-01181-0} (\bibinfo{year}{2022}).

\bibitem{NandySciData2022}
\bibinfo{author}{Nandy, A.}, \bibinfo{author}{Duan, C.} \& \bibinfo{author}{Kulik, H.~J.}
\newblock \bibinfo{journal}{\bibinfo{title}{Machine learning meets with metal organic frameworks for gas storage and separation}}.
\newblock {\emph{\JournalTitle{Journal of Chemical Information and Modeling}}} \textbf{\bibinfo{volume}{61}}, \bibinfo{pages}{2131--2146}, \doiprefix\url{10.1021/acs.jcim.1c00191} (\bibinfo{year}{2021}).

\bibitem{Terrones2024}
\bibinfo{author}{Terrones, G.~G.}, \bibinfo{author}{Huang, S.-P.}, \bibinfo{author}{Rivera, M.~P.}, \bibinfo{author}{Yue, S.}, \bibinfo{author}{Hernandez, A.} \& \bibinfo{author}{Kulik, H.~J.}
\newblock \bibinfo{journal}{\bibinfo{title}{Metal--organic framework stability in water and harsh environments from data-driven models trained on the diverse ws24 data set}}.
\newblock {\emph{\JournalTitle{Journal of the American Chemical Society}}} \textbf{\bibinfo{volume}{146}}, \bibinfo{pages}{20333--20348}, \doiprefix\url{10.1021/jacs.4c05879} (\bibinfo{year}{2024}).

\bibitem{Cao2023}
\bibinfo{author}{Cao, Z.}, \bibinfo{author}{Magar, R.}, \bibinfo{author}{Wang, Y.} \& \bibinfo{author}{Barati~Farimani, A.}
\newblock \bibinfo{journal}{\bibinfo{title}{Moformer: Self-supervised transformer model for metal--organic framework property prediction}}.
\newblock {\emph{\JournalTitle{Journal of the American Chemical Society}}} \textbf{\bibinfo{volume}{145}}, \bibinfo{pages}{2958--2967}, \doiprefix\url{10.1021/jacs.2c11420} (\bibinfo{year}{2023}).

\bibitem{Kang2023MOFTransformer}
\bibinfo{author}{Kang, Y.}, \bibinfo{author}{Park, H.}, \bibinfo{author}{Smit, B.} \& \bibinfo{author}{Kim, J.}
\newblock \bibinfo{journal}{\bibinfo{title}{A multi-modal pre-training transformer for universal transfer learning in metal--organic frameworks}}.
\newblock {\emph{\JournalTitle{Nature Machine Intelligence}}} \textbf{\bibinfo{volume}{5}}, \bibinfo{pages}{309--318}, \doiprefix\url{10.1038/s42256-023-00628-2} (\bibinfo{year}{2023}).

\bibitem{Park2024}
\bibinfo{author}{Park, J.}, \bibinfo{author}{Kim, H.}, \bibinfo{author}{Kang, Y.}, \bibinfo{author}{Lim, Y.} \& \bibinfo{author}{Kim, J.}
\newblock \bibinfo{journal}{\bibinfo{title}{From data to discovery: Recent trends of machine learning in metal--organic frameworks}}.
\newblock {\emph{\JournalTitle{JACS Au}}} \textbf{\bibinfo{volume}{4}}, \bibinfo{pages}{3727--3743}, \doiprefix\url{10.1021/jacsau.4c00618} (\bibinfo{year}{2024}).

\bibitem{Yao2021SMVAE}
\bibinfo{author}{Yao, Z.}, \bibinfo{author}{S{\'a}nchez-Lengeling, B.}, \bibinfo{author}{Bobbitt, N.~S.}, \bibinfo{author}{Bucior, B.~J.}, \bibinfo{author}{Kumar, S. G.~H.}, \bibinfo{author}{Collins, S.~P.}, \bibinfo{author}{Burns, T.}, \bibinfo{author}{Woo, T.~K.}, \bibinfo{author}{Farha, O.~K.}, \bibinfo{author}{Snurr, R.~Q.} \& \bibinfo{author}{Aspuru-Guzik, A.}
\newblock \bibinfo{journal}{\bibinfo{title}{Inverse design of nanoporous crystalline reticular materials with deep generative models}}.
\newblock {\emph{\JournalTitle{Nature Machine Intelligence}}} \textbf{\bibinfo{volume}{3}}, \bibinfo{pages}{76--86}, \doiprefix\url{10.1038/s42256-020-00271-1} (\bibinfo{year}{2021}).

\bibitem{Kim2020GAN}
\bibinfo{author}{Kim, B.}, \bibinfo{author}{Lee, S.} \& \bibinfo{author}{Kim, J.}
\newblock \bibinfo{journal}{\bibinfo{title}{Inverse design of porous materials using artificial neural networks}}.
\newblock {\emph{\JournalTitle{Science Advances}}} \textbf{\bibinfo{volume}{6}}, \bibinfo{pages}{eaax9324}, \doiprefix\url{10.1126/sciadv.aax9324} (\bibinfo{year}{2020}).
\newblock \eprint{https://www.science.org/doi/pdf/10.1126/sciadv.aax9324}.

\bibitem{Kang2024ChatMOF}
\bibinfo{author}{Kang, Y.} \& \bibinfo{author}{Kim, J.}
\newblock \bibinfo{journal}{\bibinfo{title}{Chatmof: an artificial intelligence system for predicting and generating metal-organic frameworks using large language models}}.
\newblock {\emph{\JournalTitle{Nature Communications}}} \textbf{\bibinfo{volume}{15}}, \bibinfo{pages}{4705}, \doiprefix\url{10.1038/s41467-024-48998-4} (\bibinfo{year}{2024}).

\bibitem{Yaghi2023AugJACS}
\bibinfo{author}{Zheng, Z.}, \bibinfo{author}{Zhang, O.}, \bibinfo{author}{Borgs, C.}, \bibinfo{author}{Chayes, J.~T.} \& \bibinfo{author}{Yaghi, O.~M.}
\newblock \bibinfo{journal}{\bibinfo{title}{{ChatGPT Chemistry Assistant for Text Mining and the Prediction of MOF Synthesis}}}.
\newblock {\emph{\JournalTitle{J. Am. Chem. Soc.}}} \textbf{\bibinfo{volume}{145}}, \bibinfo{pages}{18048--18062}, \doiprefix\url{10.1021/jacs.3c05819} (\bibinfo{year}{2023}).

\bibitem{Yaghi2023JACS}
\bibinfo{author}{Zheng, Z.}, \bibinfo{author}{Alawadhi, A.~H.}, \bibinfo{author}{Chheda, S.}, \bibinfo{author}{Neumann, S.~E.}, \bibinfo{author}{Rampal, N.}, \bibinfo{author}{Liu, S.}, \bibinfo{author}{Nguyen, H.~L.}, \bibinfo{author}{Lin, Y.-h.}, \bibinfo{author}{Rong, Z.}, \bibinfo{author}{Siepmann, J.~I.}, \bibinfo{author}{Gagliardi, L.}, \bibinfo{author}{Anandkumar, A.}, \bibinfo{author}{Borgs, C.}, \bibinfo{author}{Chayes, J.~T.} \& \bibinfo{author}{Yaghi, O.~M.}
\newblock \bibinfo{journal}{\bibinfo{title}{Shaping the water-harvesting behavior of metal--organic frameworks aided by fine-tuned gpt models}}.
\newblock {\emph{\JournalTitle{Journal of the American Chemical Society}}} \textbf{\bibinfo{volume}{145}}, \bibinfo{pages}{28284--28295}, \doiprefix\url{10.1021/jacs.3c12086} (\bibinfo{year}{2023}).

\bibitem{YaghiReview2025}
\bibinfo{author}{Zheng, Z.}, \bibinfo{author}{Rampal, N.}, \bibinfo{author}{Inizan, T.~J.}, \bibinfo{author}{Borgs, C.}, \bibinfo{author}{Chayes, J.~T.} \& \bibinfo{author}{Yaghi, O.~M.}
\newblock \bibinfo{journal}{\bibinfo{title}{Large language models for reticular chemistry}}.
\newblock {\emph{\JournalTitle{Nature Reviews Materials}}} \doiprefix\url{10.1038/s41578-025-00772-8} (\bibinfo{year}{2025}).

\bibitem{ddpm}
\bibinfo{author}{Ho, J.}, \bibinfo{author}{Jain, A.} \& \bibinfo{author}{Abbeel, P.}
\newblock \bibinfo{title}{Denoising diffusion probabilistic models}.
\newblock In \bibinfo{editor}{Larochelle, H.}, \bibinfo{editor}{Ranzato, M.}, \bibinfo{editor}{Hadsell, R.}, \bibinfo{editor}{Balcan, M.} \& \bibinfo{editor}{Lin, H.} (eds.) \emph{\bibinfo{booktitle}{Advances in Neural Information Processing Systems}}, vol.~\bibinfo{volume}{33}, \bibinfo{pages}{6840--6851} (\bibinfo{publisher}{Curran Associates, Inc.}, \bibinfo{year}{2020}).

\bibitem{iddpm}
\bibinfo{author}{Nichol, A.~Q.} \& \bibinfo{author}{Dhariwal, P.}
\newblock \bibinfo{title}{Improved denoising diffusion probabilistic models}.
\newblock In \emph{\bibinfo{booktitle}{Proceedings of the 38th International Conference on Machine Learning}}, \bibinfo{pages}{8162--8171} (\bibinfo{year}{2021}).

\bibitem{EDM}
\bibinfo{author}{Hoogeboom, E.}, \bibinfo{author}{Satorras, V.~G.}, \bibinfo{author}{Vignac, C.} \& \bibinfo{author}{Welling, M.}
\newblock \bibinfo{title}{Equivariant diffusion for molecule generation in 3{D}}.
\newblock In \emph{\bibinfo{booktitle}{Proceedings of the 39th International Conference on Machine Learning}}, \bibinfo{pages}{8867--8887} (\bibinfo{year}{2022}).

\bibitem{OAReactDiff}
\bibinfo{author}{Duan, C.}, \bibinfo{author}{Du, Y.}, \bibinfo{author}{Jia, H.} \& \bibinfo{author}{Kulik, H.~J.}
\newblock \bibinfo{journal}{\bibinfo{title}{Accurate transition state generation with an object-aware equivariant elementary reaction diffusion model}}.
\newblock {\emph{\JournalTitle{Nature Computational Science}}} \textbf{\bibinfo{volume}{3}}, \bibinfo{pages}{1045--1055}, \doiprefix\url{10.1038/s43588-023-00563-7} (\bibinfo{year}{2023}).

\bibitem{duan2024reactot}
\bibinfo{author}{Duan, C.}, \bibinfo{author}{Liu, G.-H.}, \bibinfo{author}{Du, Y.}, \bibinfo{author}{Chen, T.}, \bibinfo{author}{Zhao, Q.}, \bibinfo{author}{Jia, H.}, \bibinfo{author}{Gomes, C.~P.}, \bibinfo{author}{Theodorou, E.~A.} \& \bibinfo{author}{Kulik, H.~J.}
\newblock \bibinfo{journal}{\bibinfo{title}{Optimal transport for generating transition states in chemical reactions}}.
\newblock {\emph{\JournalTitle{Nature Machine Intelligence}}} \textbf{\bibinfo{volume}{7}}, \bibinfo{pages}{615--626}, \doiprefix\url{10.1038/s42256-025-01010-0} (\bibinfo{year}{2025}).

\bibitem{fx2023MOFDiff}
\bibinfo{author}{Fu, X.}, \bibinfo{author}{Xie, T.}, \bibinfo{author}{Rosen, A.~S.}, \bibinfo{author}{Jaakkola, T.} \& \bibinfo{author}{Smith, J.}
\newblock \bibinfo{title}{Mofdiff: Coarse-grained diffusion for metal-organic framework design} (\bibinfo{year}{2023}).
\newblock \eprint{2310.10732}.

\bibitem{kim2024mofflow}
\bibinfo{author}{Kim, N.}, \bibinfo{author}{Kim, S.}, \bibinfo{author}{Kim, M.}, \bibinfo{author}{Park, J.} \& \bibinfo{author}{Ahn, S.}
\newblock \bibinfo{title}{Mofflow: Flow matching for structure prediction of metal-organic frameworks} (\bibinfo{year}{2024}).
\newblock \eprint{2410.17270}.

\bibitem{joshi2025ADiT}
\bibinfo{author}{Joshi, C.~K.}, \bibinfo{author}{Fu, X.}, \bibinfo{author}{Liao, Y.-L.}, \bibinfo{author}{Gharakhanyan, V.}, \bibinfo{author}{Miller, B.~K.}, \bibinfo{author}{Sriram, A.} \& \bibinfo{author}{Ulissi, Z.~W.}
\newblock \bibinfo{title}{All-atom diffusion transformers: Unified generative modelling of molecules and materials} (\bibinfo{year}{2025}).
\newblock \eprint{2503.03965}.

\bibitem{inizan2025agenticMOF}
\bibinfo{author}{Inizan, T.~J.}, \bibinfo{author}{Yang, S.}, \bibinfo{author}{Kaplan, A.}, \bibinfo{author}{hsu Lin, Y.}, \bibinfo{author}{Yin, J.}, \bibinfo{author}{Mirzaei, S.}, \bibinfo{author}{Abdelgaid, M.}, \bibinfo{author}{Alawadhi, A.~H.}, \bibinfo{author}{Cho, K.}, \bibinfo{author}{Zheng, Z.}, \bibinfo{author}{Cubuk, E.~D.}, \bibinfo{author}{Borgs, C.}, \bibinfo{author}{Chayes, J.~T.}, \bibinfo{author}{Persson, K.~A.} \& \bibinfo{author}{Yaghi, O.~M.}
\newblock \bibinfo{title}{System of agentic ai for the discovery of metal-organic frameworks} (\bibinfo{year}{2025}).
\newblock \eprint{2504.14110}.

\bibitem{diffusion2015}
\bibinfo{author}{Sohl-Dickstein, J.}, \bibinfo{author}{Weiss, E.}, \bibinfo{author}{Maheswaranathan, N.} \& \bibinfo{author}{Ganguli, S.}
\newblock \bibinfo{title}{Deep unsupervised learning using nonequilibrium thermodynamics}.
\newblock In \emph{\bibinfo{booktitle}{International Conference on Machine Learning}}, \bibinfo{pages}{2256--2265} (\bibinfo{year}{2015}).

\bibitem{scoresde}
\bibinfo{author}{Song, Y.}, \bibinfo{author}{Sohl-Dickstein, J.}, \bibinfo{author}{Kingma, D.~P.}, \bibinfo{author}{Kumar, A.}, \bibinfo{author}{Ermon, S.} \& \bibinfo{author}{Poole, B.}
\newblock \bibinfo{title}{Score-based generative modeling through stochastic differential equations}.
\newblock In \emph{\bibinfo{booktitle}{International Conference on Learning Representations}} (\bibinfo{year}{2020}).

\bibitem{pormake}
\bibinfo{author}{Lee, S.}, \bibinfo{author}{Kim, B.}, \bibinfo{author}{Cho, H.}, \bibinfo{author}{Lee, H.}, \bibinfo{author}{Lee, S.~Y.}, \bibinfo{author}{Cho, E.~S.} \& \bibinfo{author}{Kim, J.}
\newblock \bibinfo{journal}{\bibinfo{title}{Computational screening of trillions of metal--organic frameworks for high-performance methane storage}}.
\newblock {\emph{\JournalTitle{ACS Applied Materials {\&} Interfaces}}} \textbf{\bibinfo{volume}{13}}, \bibinfo{pages}{23647--23654}, \doiprefix\url{10.1021/acsami.1c02471} (\bibinfo{year}{2021}).

\bibitem{Bucior2019}
\bibinfo{author}{Bucior, B.~J.}, \bibinfo{author}{Rosen, A.~S.}, \bibinfo{author}{Haranczyk, M.}, \bibinfo{author}{Yao, Z.}, \bibinfo{author}{Ziebel, M.~E.}, \bibinfo{author}{Farha, O.~K.}, \bibinfo{author}{Hupp, J.~T.}, \bibinfo{author}{Siepmann, J.~I.}, \bibinfo{author}{Aspuru-Guzik, A.} \& \bibinfo{author}{Snurr, R.~Q.}
\newblock \bibinfo{journal}{\bibinfo{title}{Identification schemes for metal–organic frameworks to enable rapid search and cheminformatics analysis}}.
\newblock {\emph{\JournalTitle{Crystal Growth \& Design}}} \textbf{\bibinfo{volume}{19}}, \bibinfo{pages}{6682–6697}, \doiprefix\url{10.1021/acs.cgd.9b01050} (\bibinfo{year}{2019}).

\bibitem{YaghiChemRevNets2014}
\bibinfo{author}{Li, M.}, \bibinfo{author}{Li, D.}, \bibinfo{author}{O'Keeffe, M.} \& \bibinfo{author}{Yaghi, O.}
\newblock \bibinfo{journal}{\bibinfo{title}{Topological analysis of metal–organic frameworks with polytopic linkers and/or multiple building units and the minimal transitivity principle}}.
\newblock {\emph{\JournalTitle{Chemical Reviews}}} \bibinfo{pages}{1343–1370}, \doiprefix\url{10.1021/cr400392k} (\bibinfo{year}{2014}).

\bibitem{EddaoudiChemRevNets2020}
\bibinfo{author}{Chen, Z.}, \bibinfo{author}{Jiang, H.}, \bibinfo{author}{Li, M.}, \bibinfo{author}{O'Keeffe, M.} \& \bibinfo{author}{Eddaoudi, M.}
\newblock \bibinfo{journal}{\bibinfo{title}{Reticular chemistry 3.2: Typical minimal edge-transitive derived and related nets for the design and synthesis of metal–organic frameworks}}.
\newblock {\emph{\JournalTitle{Chemical Reviews}}} \bibinfo{pages}{8039–8065}, \doiprefix\url{10.1021/acs.chemrev.9b00648} (\bibinfo{year}{2020}).

\bibitem{leftnet}
\bibinfo{author}{Du, W.}, \bibinfo{author}{Du, Y.}, \bibinfo{author}{Wang, L.}, \bibinfo{author}{Feng, D.}, \bibinfo{author}{Wang, G.}, \bibinfo{author}{Ji, S.}, \bibinfo{author}{Gomes, C.} \& \bibinfo{author}{Ma, Z.-M.}
\newblock \bibinfo{journal}{\bibinfo{title}{A new perspective on building efficient and expressive 3{D} equivariant graph neural networks}}.
\newblock {\emph{\JournalTitle{arXiv:2304.04757}}}  (\bibinfo{year}{2023}).

\bibitem{JanetJPCA2017}
\bibinfo{author}{Janet, J.} \& \bibinfo{author}{Kulik, H.~J.}
\newblock \bibinfo{journal}{\bibinfo{title}{Resolving transition metal chemical space: Feature selection for machine learning and structure–property relationships}}.
\newblock {\emph{\JournalTitle{The Journal of Physical Chemistry A}}} \bibinfo{pages}{8939–8954}, \doiprefix\url{10.1021/acs.jpca.7b08750} (\bibinfo{year}{2017}).

\bibitem{2013Influential}
\bibinfo{author}{Yang, S.~Y.}, \bibinfo{author}{Yuan, H.~B.}, \bibinfo{author}{Xu, X.~B.} \& \bibinfo{author}{Huang, R.~B.}
\newblock \bibinfo{journal}{\bibinfo{title}{Influential factors on assembly of first-row transition metal coordination polymers}}.
\newblock {\emph{\JournalTitle{Inorganica Chimica Acta}}} \textbf{\bibinfo{volume}{403}}, \bibinfo{pages}{53--62} (\bibinfo{year}{2013}).

\bibitem{2007Tri}
\bibinfo{author}{Luisi, B.~S.}, \bibinfo{author}{Ma, Z.} \& \bibinfo{author}{Moulton, B.}
\newblock \bibinfo{journal}{\bibinfo{title}{Tri-metal secondary building units: Toward the design of thermally robust crystalline coordination polymers}}.
\newblock {\emph{\JournalTitle{Journal of Chemical Crystallography}}} \textbf{\bibinfo{volume}{37}}, \bibinfo{pages}{743--747} (\bibinfo{year}{2007}).

\bibitem{clf_free_diffusion}
\bibinfo{author}{Ho, J.} \& \bibinfo{author}{Salimans, T.}
\newblock \bibinfo{title}{Classifier-free diffusion guidance} (\bibinfo{year}{2022}).
\newblock \eprint{2207.12598}.

\bibitem{equivariance}
\bibinfo{author}{Serre, J.-P.} \emph{et~al.}
\newblock \emph{\bibinfo{title}{Linear representations of finite groups}}, vol.~\bibinfo{volume}{42} (\bibinfo{publisher}{Springer}, \bibinfo{year}{1977}).

\bibitem{gdlbook}
\bibinfo{author}{Bronstein, M.~M.}, \bibinfo{author}{Bruna, J.}, \bibinfo{author}{Cohen, T.} \& \bibinfo{author}{Veli{\v{c}}kovi{\'c}, P.}
\newblock \bibinfo{journal}{\bibinfo{title}{Geometric deep learning: Grids, groups, graphs, geodesics, and gauges}}.
\newblock {\emph{\JournalTitle{arXiv:2104.13478}}}  (\bibinfo{year}{2021}).

\bibitem{campbell2022continuous}
\bibinfo{author}{Campbell, A.}, \bibinfo{author}{Benton, J.}, \bibinfo{author}{De~Bortoli, V.}, \bibinfo{author}{Rainforth, T.}, \bibinfo{author}{Deligiannidis, G.} \& \bibinfo{author}{Doucet, A.}
\newblock \bibinfo{journal}{\bibinfo{title}{A continuous time framework for discrete denoising models}}.
\newblock {\emph{\JournalTitle{Advances in Neural Information Processing Systems}}} \textbf{\bibinfo{volume}{35}}, \bibinfo{pages}{28266--28279} (\bibinfo{year}{2022}).

\bibitem{Repaint}
\bibinfo{author}{Lugmayr, A.}, \bibinfo{author}{Danelljan, M.}, \bibinfo{author}{Romero, A.}, \bibinfo{author}{Yu, F.}, \bibinfo{author}{Timofte, R.} \& \bibinfo{author}{Van~Gool, L.}
\newblock \bibinfo{title}{Repaint: Inpainting using denoising diffusion probabilistic models}.
\newblock In \emph{\bibinfo{booktitle}{2022 IEEE/CVF Conference on Computer Vision and Pattern Recognition (CVPR)}}, \doiprefix\url{10.1109/CVPR52688.2022.01117} (\bibinfo{year}{2022}).

\bibitem{LeeACSAMI2021}
\bibinfo{author}{Lee, S.}, \bibinfo{author}{Kim, B.}, \bibinfo{author}{Hyun, C.}, \bibinfo{author}{Lee, H.}, \bibinfo{author}{Yunmi~Lee, S.}, \bibinfo{author}{Seon~Cho, E.} \& \bibinfo{author}{Kim, J.}
\newblock \bibinfo{journal}{\bibinfo{title}{Computational screening of trillions of metal–organic frameworks for high-performance methane storage}}.
\newblock {\emph{\JournalTitle{ACS Applied Materials \& Interfaces}}} \textbf{\bibinfo{volume}{13}}, \bibinfo{pages}{23647--23654}, \doiprefix\url{10.1021/acsami.1c02471} (\bibinfo{year}{2021}).

\end{thebibliography}
